\documentclass{article}

\usepackage{arxiv}

\usepackage[utf8]{inputenc} 
\usepackage[slovak,english]{babel}
\usepackage{hyperref}       
\usepackage{url}            
\usepackage{booktabs}       
\usepackage{nicefrac}       
\usepackage{microtype}      
\usepackage{graphicx}
\usepackage{natbib}

\usepackage{algorithmic}
\usepackage{amssymb,amsmath,xcolor,amsfonts}
\usepackage[normalem]{ulem}
\usepackage{graphicx}
\usepackage{wrapfig} 
\usepackage{url}
\usepackage{textcomp}
\usepackage{multicol}
\usepackage{array}
\usepackage{mathtools}

\usepackage[shortlabels]{enumitem}
\usepackage{mdframed}
\title{Trajectory-based Safety of Monotone Systems:\\Verification and Control Synthesis}

\newcommand{\eps}{\varepsilon}

\newcommand{\abs}[1]{{\left\lvert{#1}\right\rvert}}

\newcommand{\mc}[1]{\mathcal{{#1}}}
\newcommand{\mbf}[1]{\mathbf{{#1}}}
\newcommand{\mbb}[1]{\mathbb{{#1}}}

\DeclareSymbolFont{bbold}{U}{bbold}{m}{n}
\DeclareSymbolFontAlphabet{\mathbbold}{bbold}
\newcommand{\vect}[1]{\mathbbold{#1}}

\newcommand{\mU}{\mathcal{U}}
\newcommand{\mW}{\mathcal{W}}
\newcommand{\mV}{\mathcal{V}}
\newcommand{\mX}{\mathcal{X}}

\newcommand{\mS}{\mathfrak{S}}

\newcommand{\diag}[1]{\text{diag}\left(#1\right)}

\renewcommand{\Re}{\mathbb{R}}
\newcommand{\Rn}{\mathbb{R}^n}

\newcommand{\NN}{\mathbb{N}}

\newcommand{\norm}[1]{\left\| {#1}\right\|_{\infty}}

\newcommand{\diss}{\mbb{P}}
\newcommand{\adiss}{\mbb{Q}}

\newcommand{\barrier}{\mathbb{B}}
\newcommand{\testx}{\tilde{\mathbf{x}}}
\newcommand{\testu}{\tilde{\mathbf{u}}}
\newcommand{\testw}{\tilde{\mathbf{w}}}
\newcommand{\hatx}{\hat{\mathbf{x}}}

\newcommand{\hatw}{\hat{\mathbf{w}}}
\newcommand{\ctrl}{\tilde{\pi}}
\newcommand{\lipconst}[1]{\mathcal{L}^{#1}}

\newtheorem{assumption}{Assumption}
\newtheorem{definition}{Definition}

\newtheorem{theorem}{Theorem}

\newtheorem{proposition}{Proposition}
\newtheorem{remark}{Remark}

\newtheorem{example}{Example}
\newtheorem{problem}{Problem}

\author{{\hspace{1mm}Felipe Galarza-Jimenez}\\
	Department of Computer Science\\
	University of Colorado at Boulder\\
	\texttt{felipe.galarzajimenez@colorado.edu} \\
	\And
	{\hspace{1mm}Majid Zamani} \\
	Department of Computer Science\\
	University of Colorado at Boulder\\
	\texttt{majid.zamani@colorado.edu} \\
    \And
	{\hspace{1mm}Saber Jafarpour} \\
	Department of Computer Science\\
	University of Colorado at Boulder\\
	\texttt{saber.jafarpour@colorado.edu} \\
}

\date{}

\renewcommand{\shorttitle}{Trajectory-based Safety of Monotone Systems}

\hypersetup{
pdftitle={Trajectory-based Safety of Monotone Systems:Verification and Control Synthesis},
pdfsubject={data-driven control},
pdfauthor={Felipe Galarza-Jimenez, Majid Zamani, Saber Jafarpour},
pdfkeywords={Safety Verification, Control Synthesis, Barrier Certificates, Monotone Systems},
}

\begin{document}
\maketitle

\begin{abstract}
This paper presents a novel data-driven framework for the robust safety verification and safe control synthesis of unknown monotone discrete-time systems.
While existing data-driven safety analysis approaches are often either heuristic in nature or require large amounts of data to provide rigorous guarantees, we leverage the structural property of \emph{monotonicity} to significantly reduce data requirements while still ensuring formal safety guarantees.
Our approach is built upon a new class of certificates called \emph{dominance functions}, constructed directly from collected system trajectories, which themselves need not be safe.
By exploiting the monotone structure of the dynamics, we show that dominance functions are (i) \emph{dissipative}, meaning that they decrease monotonically along system trajectories, and (ii) sufficiently \emph{expressive} to characterize safety certificates for monotone systems.
Together, these properties establish dominance functions as principled building blocks for the systematic construction of formal safety certificates directly from trajectory data.
For both robust safety verification and safe control synthesis, we develop an efficient sampling-based optimization framework that searches for safety certificates represented as linear combinations of dominance functions constructed from collected trajectories.
We validate our data-driven framework on two monotone systems by successfully deriving safety certificates from a small number of trajectories. 
\end{abstract}

\keywords{Safety Verification \and Control Synthesis \and Barrier Certificates \and Monotone Systems.}

\section{Introduction}
The rapid deployment of autonomous systems in safety-critical domains is reshaping sectors such as transportation, aerospace, manufacturing, and energy.
Providing rigorous safety guarantees for these systems is essential, as failures or unsafe behavior can lead to catastrophic consequences.
For many autonomous systems, a central challenge to ensuring safety is the presence of unknown components or dynamics for which accurate models are unavailable and operational data is the only available source for characterizing their behavior.
Examples include learning-based modules that are inherently data-driven, perception systems whose behavior cannot be described through first-principles models, and lookup-table controllers for which deriving an explicit analytical representation is highly complex.
Traditional frameworks for providing safety guarantees for dynamical systems typically rely on full knowledge of the system dynamics.
Examples include abstraction-based techniques~\cite{tabuada2009verification,zamani2011symbolic}, which use simplified finite-state models for verification, inductive methods that construct certificates, such as barrier functions, to prove safety through invariance arguments~\cite{Prajna2004,Ames2019}, and reachability approaches that over-approximate the reachable set of the system~\cite{MA-GF-AG:21,XC-EA-SS:13,SB-MC-SH-CJT:17}.

In systems where the full dynamics or some of their components are unknown, data-driven safety assurance techniques have recently emerged as a practical alternative to traditional model-based verification for providing rigorous safety guarantees~\cite{wabersich2023data}.
These techniques are generally categorized into two main approaches.
The first category, referred to as \textit{indirect} methods, relies on collected data to construct a surrogate model, which is subsequently used to derive a safety certificate.
Examples include~\cite{wang2018safe}, which uses Gaussian processes to learn models from data in partially unknown environments, and~\cite{black2023safe,zhou2025learning}, which estimate system models via Koopman operator approximations from data and then construct safety or stability certificates.
Despite several important developments, the effectiveness of indirect methods often depends critically on the accuracy of the surrogate model and may lead to overly conservative safety certificates, even in the presence of relatively small modeling errors.

The second category, referred to as \textit{direct} methods, constructs safety certificates directly from system data without explicitly identifying a surrogate model.
Depending on how the data is collected, direct methods can be broadly categorized into \textit{trajectory-based} approaches, which rely on observed system trajectories, and \textit{simulator-based} approaches, which use direct simulator queries over different regions of the state space.
Trajectory-based methods construct safety certificates directly from observed trajectories~\cite{de2019formulas}, but they are often limited to certain classes of systems or restrict the certificate search to particular set representations~\cite{luppi2024data}.
Moreover, they typically require sufficiently rich or persistently exciting data to accurately capture the system behavior, which may be difficult or unsafe to obtain in safety-critical settings~\cite{de2019formulas}.
Simulator-based methods, on the other hand, infer safety properties by querying the system over discretizations of the state space~\cite{nejati2022data}.
Although more broadly applicable, simulator-based methods are typically highly data-intensive, as obtaining rigorous guarantees requires sufficiently dense coverage of the state space---often growing exponentially with the system dimension---and commonly relies on Lipschitz bounds of the system to extend guarantees between sampled data~\cite{wang2019data,anand2023formally,robey2020learning}.

These limitations motivate the study of structural system properties that enable formal verification guarantees to be established from limited data.
In particular, monotonicity---a common property in many dynamical systems---has recently attracted significant attention for its potential to alleviate these challenges.
By preserving the standard partial order along system trajectories, monotone systems admit more efficient analysis, verification, and certification procedures~\cite{angeli2003monotone,dirr2015separable,SJ-AD-FB:20r}.
Recent studies have shown that monotonicity enables efficient stability certification~\cite{AR:15,sootla2016construction,coogan2019contractive,SJ-SC:25}, formal reachability analysis~\cite{SC:20}, and control synthesis~\cite{sinyakov2019reachability}.
Furthermore, the computation of robust and controlled invariant sets for monotone and mixed-monotone systems has been extensively studied, both using system models~\cite{sadraddini2016safety,coogan2020mixed,abate2022robustly,ivanova2022lazy,de2007monotonicity} and directly from data~\cite{makdesi2023data,sinyakov2020abstraction,MA-AS:24}.
However, despite these developments, the use of monotonicity for the data-driven verification and synthesis of barrier certificates remains largely unexplored, with the exception of~\cite{alavi2025neural}.

\paragraph*{Contributions}
In this paper, we propose a novel data-driven framework for robust safety verification and safe control synthesis of monotone discrete-time systems directly from trajectory data, without requiring explicit knowledge of the underlying system dynamics.
First, we introduce a new class of data-driven functions, called \emph{dominance functions}, with formulations tailored to two distinct interpretations of system inputs.
In the first setting, inputs are treated as unknown disturbances, for which we construct robustified dominance functions using Lipschitz bounds on the dynamics.
In the second setting, inputs correspond to feedback control policies, where dominance functions are extended to explicitly incorporate input sequences for safe control synthesis.
Second, by leveraging the monotone structure of the dynamics, we show that dominance functions are dissipative, meaning that they decrease monotonically along system trajectories.
This dissipation property reveals a fundamental connection between dominance functions and inductive safety reasoning via barrier certificates, enabling system safety properties to be inferred directly from data.
Building on this connection, we show that, in the disturbance-free setting, dominance functions are sufficiently expressive to construct robust barrier certificates for monotone systems.
This result motivates their use as principled building blocks for constructing formal safety certificates of monotone systems from trajectory data.

For both robust safety verification and safe control synthesis, we formulate the search for safety certificates as an optimization problem over linear combinations of dominance functions constructed from collected trajectories.
By leveraging the monotone structure of these dominance functions, we develop an efficient sampling-based approach for solving the resulting optimization problem, thereby enabling scalable and data-efficient safety certification.
Finally, we validate the proposed framework on safety verification problems in population dynamics and safe control synthesis tasks in traffic networks, demonstrating the construction of formal safety certificates from only a small number of collected trajectories.

Compared to our conference version~\cite{galarza2025trajectory}, this paper provides a substantially more complete treatment of dominance functions by extending the framework to systems with two classes of inputs: disturbances and control actions.
This extension enables the study of a broader range of problems, including robust safety verification and safe control synthesis.
Moreover, for monotone systems without disturbances, we show that the class of dominance functions is sufficiently expressive to characterize system safety.
Finally, the paper includes all proofs omitted from the conference version.

\section{Notation and Preliminaries}\label{sec:preliminaries}
We denote by $\NN$ and $\Re$ the set of non-negative integers including $\{+\infty\}$ and the set of real numbers, respectively. Given $a \in \NN$ ($a \in \Re$), we use $\NN_{\geq a}$ $(\text{resp. }\Re_{\geq a})$ to denote all values in $\NN$ $(\text{resp. }\Re)$ greater than or equal to $a$. For $a,b \in \Re$, $[a,b]$, $]a,b[$, $[a,b[$, and $]a,b]$ denote closed, open, and half-open interval in $\Re$.
Likewise, $[a;b]$, $]a;b[$, $[a;b[$ and $]a;b]$ denote closed, open, and half-open sets in $\NN$.
Given sets $X_1, \ldots, X_n$, for some $n \in \NN_{\geq 1}$, we denote their Cartesian product by $X_1 \times X_2 \times \ldots \times X_n$.
We define $\vect{1}_n \in \Rn$ as the vector in $\Re^n$ whose components are all equal to $1$.
We denote the infinity norm of an element of a vector space by $\norm{\cdot}$.

Given sets $A$ and $B$, we represent the set difference as $A \backslash B:=\{a \in A ~|~ a \notin  B\}$ and $\abs{A}$ denotes the cardinality of the set $A$. We use $f:A \rightarrow B$ to denote a function from $A$ to $B$. We use $f(A)$ to denote the set $\{f(a) \in B \mid \text{ for all } a \in A \}$. 

Given $x,y \in \Rn$, we write $x\le y$ if $x_j\le y_j$, for every $j\in [1;n]$. 
We denote the interval $\{z\in \Re^n\mid x\le z\le y\}$ by $[x,y]$.
Given a set $X \subset \Rn$, for any subset $Y \subseteq X$, we define its \textit{lower (upper) closure} as $[Y]_{\downarrow}:= \bigcup_{y \in Y}\{x \in X \mid x \le y\}$ ($[Y]_{\uparrow}:= \bigcup_{y \in Y}\{x \in X \mid y \le x\}$). Moreover, we say that $Y \subset X$ is \textit{lower (upper) } closed if $Y = [Y]_{\downarrow} (Y = [Y]_{\uparrow})$. 
%
Given a map $f:\Re^n\to \Re^m$, the \emph{inclusion function} \cite{jaulin2001interval} of $f$ is a map $\mathsf{F} = \left(\underline{\mathsf{F}},\overline{\mathsf{F}}\right):\Re^{2n}\to \Re^{2m}$ such that 
\begin{align*}
  \underline{\mathsf{F}}(\underline{x},\overline{x}) \le f(x) \le \overline{\mathsf{F}}(\underline{x},\overline{x}),\qquad\mbox{ for all }x\in [\underline{x},\overline{x}]. 
\end{align*}

\subsection{Discrete-time Monotone Systems}\label{sec:monotone}
In this part, we introduce discrete-time monotone systems and characterize their properties.

\begin{definition}[Discrete-time System \cite{smith1995monotone}]\label{def:dtsys}
A discrete-time dynamical system $\mS$ is a tuple $\mS = (\mX,\mX_0,\mV,f)$, where $\mX\subseteq \Rn$ is the state set, $\mX_0 \subseteq \mX$ is the initial state set, $\mV \subseteq \Re^v$ is the input set, and $f: \mX \times \mV \rightarrow \mX$ is a jointly continuous state transition map. The evolution of the states of a discrete-time system $\mS=(\mX,\mX_0,\mV,f)$ is given by
\begin{align}\label{eq:dtsys}
        \mbf{x}(t+1) = f(\mbf{x}(t),\mbf{v}(t)),\qquad \mbox{ for all }t \in \NN.
\end{align} 
\end{definition}

Given $x_0\in \mX_0$ and an input sequence $\mbf{v} = (\mbf{v}(0),\ldots,\mbf{v}(t),\ldots)\in \mV^{\omega}$, the trajectory of the discrete-time system $\mS$ starting from $x_0$ is defined by $\mbf{x}:=(\mbf{x}(0),\ldots,\mbf{x}(t),\ldots)\in \mX^{\omega}$, with the convention that $\textbf{x}(0) = x_0$. For a given $T \in \NN$, the $T$-truncated trajectory of the discrete-time system $\mS$ starting from $x_0$ is defined by $\mbf{x}^{T}:=(\mbf{x}(0),\ldots,\mbf{x}(T))\in \mX^{[0;T]}$ with a small abuse of notation to highlight the temporal nature of the vector.

\begin{definition}[Monotone systems]
Given a discrete-time dynamical system $\mS=(\mX,\mX_0,\mV,f)$, we say $\mS$ is
\begin{enumerate}
    \item \emph{state monotone (SM)}, if for all $v\in \mV$,
    \begin{align}\label{eq:smon}
    x\le x'\implies f(x,v)\le f(x',v);
    \end{align}
    \item \emph{state and input monotone (SIM)}, if
\begin{align}\label{eq:simon}
       x \leq x', v \leq v' &\implies f(x,v) \leq f(x',v'). 
\end{align} 
\end{enumerate}
\end{definition}
\smallskip
\noindent
In this work, we focus on two notions of safety: \emph{robust safety} and \emph{controlled safety}. The distinction between these concepts arises from how we interpret the input set.
\begin{definition}[Robust Safety]\label{def:robustsafe}
 Given a discrete-time dynamical system $\mS = (\mX,\mX_0,\mW,f)$, a set of unsafe states $\mX_u\subseteq \mX$ and a set $\mathcal{S}\subseteq \mX$, then
\begin{enumerate}
    \item $\mathcal{S}$ is a \textit{robust invariant set} for $\mS$, if for every initial condition $x_0\in \mX_0$ and every disturbance sequence $\mbf{w} \in \mW^\omega$, we have $\mbf{x}(t)\in \mathcal{S}$, for every $t\in \NN$.
    \item $\mS$ is \textit{robustly safe} with respect to $\mX_u$, if there exists a robust invariant set $\mathcal{S}\subseteq \mX$ for $\mS$ such that $\mathcal{S}\cap \mX_u=\emptyset$. 
\end{enumerate}
\end{definition}

Roughly speaking, for the \emph{robust safety} of a discrete-time dynamical system $\mS = (\mX, \mX_0, \mW, f)$, we interpret the input set $\mW$ as an \emph{disturbance}. The goal is to verify that the system remains safe under all worst-case disturbance realizations. 
In this work, we leverage the concept of \textit{robust barrier certificates} to provide a function characterization of robust safety. We formalize as stated in the following definitions. 

\begin{definition}[Robust Barrier Certificate (RBC)]\label{def:robust_barrier}
    Consider a discrete-time system $\mS = (\mX,\mX_0,\mW,f)$ with an unsafe set $\mX_u \subseteq \mX$. Then, a function $\barrier: \mX \rightarrow \Re$ is called a \textit{Robust Barrier Certificate (RBC)} if:
\begin{subequations}\label{eq:rbc}
    \begin{align}
        &\barrier(x)\leq 0, &&\mbox{ for all } x \in \mX_0, \label{eq:rbc1}\\
        &\barrier(x) > 0, &&\mbox{ for all } x \in \mX_u, \label{eq:rbc2}\\
        &\barrier(f(x,w)) \leq \barrier(x), &&\mbox{ for all } w\in \mW \mbox{ and } x\in \mX. \label{eq:rbc3}
    \end{align}
\end{subequations}
\end{definition}
The existence of an RBC for $\mS$ with unsafe set $\mX_u$, grants that the $0-$sublevel set of $\barrier$, defined by $\barrier_{\leq 0}:=\{x \in \mX \mid \barrier(x) \leq 0\}$ is a \textit{robust invariant set} \cite{Prajna2004}. Then, since $\mX_u \cap \barrier_{\leq 0} = \emptyset$, the system $\mS$ is robustly safe as in Definition \ref{def:robustsafe}.

\begin{definition}[Controlled Safety]\label{def:controlledsafe}
 Given a discrete-time dynamical system $\mS = (\mX,\mX_0,\mU,f)$, a set of unsafe states $\mX_u\subseteq \mX$ and a set $\mathcal{S}\subseteq \mX$, then
 \begin{enumerate}
     \item  $\mathcal{S}$ is a \textit{controlled invariant set} for $\mS$, if for every initial condition $x_0\in \mX_0$ there exists an input sequence $\mbf{u} \in \mU^\omega$ such that $\mbf{x}(t)\in \mathcal{S}$, for every $t\in \NN$.
     \item $\mS$ is \textit{controlled safe} with respect to $\mX_u$ if there exists a controlled invariant set $\mathcal{S}$ for $\mS$ such that $\mathcal{S}\cap \mX_u=\emptyset$. 
 \end{enumerate}
\end{definition}

Roughly speaking, for the \emph{controlled safety} of a discrete-time dynamical system $\mS = (\mX,\mX_0,\mU,f)$, we treat the input set $\mU$ as the \textit{control actions} and seek to synthesize a controller that ensures the system remains safe. The notion of a \textit{control barrier certificate} has been widely used in the literature to study controlled safety of systems~\cite{Ames2019}.

\begin{definition}[Control Barrier Certificate (CBC)]\label{def:cbarrier}
    Consider a system $\mS = (\mX,\mX_0,\mU,f)$ and an unsafe set $\mX_u \subseteq \mX$. Then, a function $\barrier: \mX \rightarrow \Re$ is called a \textit{Control Barrier Certificate (CBC)} if:
\begin{subequations}\label{eq:cbarrierdef}
    \begin{align}
        &\barrier(x)\leq 0, \mbox{ for all } x \in \mX_0, \label{eq:cbc1}\\
        &\barrier(x) > 0, \mbox{ for all } x \in \mX_u, \label{eq:cbc2}\\
        &\mbox{For all } x \in \mX\backslash \mX_u, \mbox{ there is } u \in \mU\mbox{ such that:} \nonumber\\
        &\barrier(f(x,u)) \leq \barrier(x). \label{eq:cbc3}
    \end{align}
\end{subequations}
\end{definition}
\smallskip

The existence of a $CBC$ for $\mS$ with unsafe set $\mX_u$, grants that the $0-$sublevel set of $\barrier$, defined by $\barrier_{\leq 0}:=\{x \in \mX \mid \barrier(x) \leq 0\}$ is a \textit{controlled invariant set}~\cite{Ames2019}. Then, since $\mX_u \cap \barrier_{\leq 0} = \emptyset$, the system $\mS$ is controlled safe as in Definition \ref{def:controlledsafe}.
\subsection{Problem Formulation} 
In this paper, we study robust safety and controlled safety of discrete-time monotone systems in the setting where the transition map is \textit{unknown}, and only a limited collection of trajectories is available, with \textit{no access} to a system simulator. 
As the role of inputs differs fundamentally between the robust safety and controlled safety formulations, the information contained in each collected trajectory differs between the two settings.
In the robust safety setting, trajectories may evolve under arbitrary admissible disturbances; therefore, the collected data do not provide explicit information about the underlying disturbance signals. 

\begin{assumption}[Trajectory data with disturbance]\label{as:test_verification}
For a SM system $\mS = (\mX,\mX_0,\mW, f)$, we have access to $N$ collected trajectories $\testx_1,\testx_2,\ldots,\testx_N$ starting from $x_{0,1},$ $x_{0,2},\ldots,x_{0,N}\in \mX$ which are subject to \textit{unknown} disturbances $\testw_1,\testw_2,\ldots,\testw_N\in \mathcal{W}$, respectively.
\end{assumption}

In the controlled safety setting, trajectories are generated under designed control policies to accomplish a prescribed objective; as a result, the collected trajectory data includes explicit information about the applied control inputs. 

\begin{assumption}[Controlled trajectory data]\label{as:test_control}
For a SIM system $\mS=(\mX, \mX_0,\mU, f)$ and $N$ \textit{known} monotone feedback controllers $\ctrl_k:\mX \to \mU$ for all $k \in [1;N]$, we have access to $N$ collected trajectories $\testx_1,\ldots,\testx_N$ starting from $x_{0,1},\ldots, x_{0,N} \in \mX$ with the control inputs $\testu_1,\ldots,\testu_N$, respectively, where  $\testu_k(t) = \ctrl_k(\testx_k(t))$ for all $t \in \NN$.  
\end{assumption}

Our primary objective is to develop a theoretical framework for analyzing robust and controlled safety of unknown monotone systems using only a finite collection of trajectories, as described in Assumptions~\ref{as:test_verification} and~\ref{as:test_control}, respectively. We formalize these goals through the following two problems.

\begin{problem}[Robust Safety Verification]\label{prob1}
Given an SM system $\mS = (\mX,\mX_0,\mW,f)$ with an unsafe set $\mX_u \subseteq \mX$, and an unknown transition map $f$ with a finite family of collected trajectories satisfying Assumption~\ref{as:test_verification}, verify whether $\mS$ is robustly safe with respect to $\mX_u$.
\end{problem}

\begin{problem}[Safe Control Synthesis]\label{prob2}
Given a SIM system $\mS = (\mX,\mX_0,\mU,f)$ with an unsafe set $\mX_u \subseteq \mX$, and an unknown transition map $f$ with a finite family of collected trajectories satisfying Assumption~\ref{as:test_control}, design a state feedback controller $\pi: \mX \to \mU$ which makes the system \textit{controlled safe} with respect to $\mX_u$.
\end{problem}

In the following sections, we address Problem~\ref{prob1} and Problem~\ref{prob2} by introducing a new class of trajectory-based functions and leveraging the system’s monotonicity to construct barrier certificates with these functions as building blocks.

\section{Trajectory-based Robust Verification}\label{sec:verification}

In this section, we focus on Problem~\ref{prob1} and develop a framework that uses the collected trajectories of the unknown SM system $\mS$ to certify its robust safety. We start with the following assumption. 

\begin{assumption}[Known Lipschitz Bounds]\label{as:lipschitz}
 The discrete-time system $\mS = (\mX,\mX_0,\mW,f)$ has a Lipschitz continuous transition map with respect to $x$ and $w$ with \textit{known} Lipschitz constants, i.e., there exist \emph{known} constants $L_x,L_w \in \Re_{>0}$,
\begin{align*}
    &\norm{f(x,w) - f(x',w)} \leq L_x \norm{x - x'},\\
    &\norm{f(x,w) - f(x,w')} \leq L_w \norm{w - w'},
\end{align*}
\end{assumption}
\smallskip

 Combining Assumption~\ref{as:lipschitz} with Assumption~\ref{as:test_verification} allows us to bound the effect of disturbances on the system trajectory, as formalized in the following proposition.

\begin{proposition}[Trajectory Comparison]\label{thm:traj_comp}
Consider an SM discrete-time system $\mS = (\mX,\mX_0,\mW,f)$ satisfying Assumption~\ref{as:lipschitz}. Let $\testx$ and $\hatx$ be two trajectories of $\mS$ starting from the same initial condition $\testx(0) =\hatx(0)$ with different realizations of disturbances. For every $t\ge 0$,  
\begin{align*}
    - (L_w D_w\sum_{\tau=0}^{t-1}L_x^{\tau})\vect{1}_n \leq \hatx(t) - \testx(t)  \leq (L_w D_w\sum_{\tau=0}^{t-1}L_x^{\tau})\vect{1}_n.
\end{align*}
\end{proposition}

\paragraph*{Proof:} See Appendix \ref{ap:thmtraj_comp}.
\smallskip

Proposition~\ref{thm:traj_comp} quantifies how the \emph{error} between two trajectories subjected to different disturbances propagates over time. In the special case where $L_x < 1$, i.e., when $\mS$ is contractive with respect to $x$, this propagation can be uniformly bounded for any $t \in \NN$ by $L_wD_w\sum_{\tau=0}^{t-1} L_x^{\tau}\vect{1}_n \leq \frac{L_wD_w}{1 - L_x}\vect{1}_n$. From now on, for the sake of simplicity of notation, we denote 
\begin{align}\label{eq:compact}
\lipconst{t-1}:= L_w D_w\sum_{\tau=0}^{t-1}L_x^{\tau}.    
\end{align}

\subsection{Robust Dominance Functions}

In the first step, for each trajectory, we introduce two associated data-driven functions, referred to as the robust upper and lower dominance functions. These functions will serve as building blocks for constructing robust safety certificates.
\begin{definition}[Robust Upper Dominance functions]\label{def:disshatfcn} Let $\mS = (\mX, \mX_0, \mW, f)$ be an SM system satisfying Assumption~\ref{as:lipschitz}. Given a trajectory $\testx$ generated under an arbitrary disturbance and initial condition $\testx(0) \in \mX$, the \emph{robust upper dominance time} $t^{\leq} : \mX \to \Re_{\ge 0} \cup \{\infty\}$ is defined by
\begin{align}\label{eq:Phattime}
t^{\leq}(x)&:=\sup \left\{t \mid x \leq \testx(t) + \lipconst{t-1}\vect{1}_n \right\},
\end{align}
where $\lipconst{t-1}$ is as defined in~\eqref{eq:compact}. Then, for a given parameter $\alpha\in \Re_{>1}$, the \textit{robust upper dominance function} $\diss^{\testx}:\mathcal{X}\to \Re$ is defined by
\begin{align}\label{eq:disshatfcn}
 \diss^{\testx} (x)&:= \begin{cases}
     \tfrac{1}{t^{\leq}(x)+1} & t^{\leq}(x) < \infty,\\
     0 & t^{\leq}(x) = \infty,\\
     \alpha & t^{\leq}(x) = \emptyset.
 \end{cases}
\end{align}    
\end{definition}
\smallskip

\begin{figure}
        \centering
        \includegraphics[width=1\linewidth]{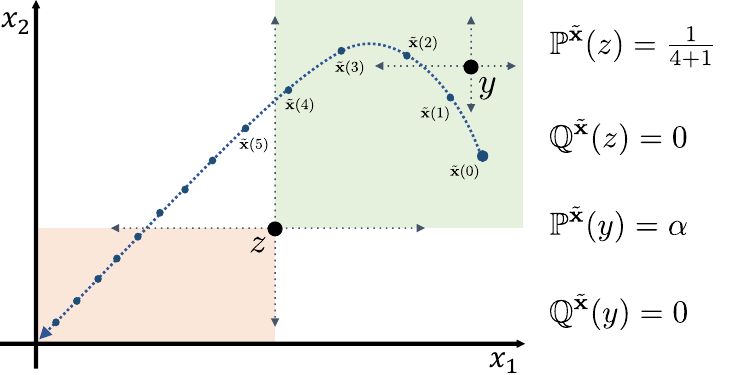}
        \caption{Illustration of the robust upper and robust lower dominance times $t^{\le}$ and $t^{\ge}$ and the robust upper and robust lower dominance functions $\diss^{\testx}$ and $\adiss^{\testx}$ associated to a given trajectory $\testx$.   These functions are evaluated at state points $z$ and $y$ shown above. For point $z$, the robust upper dominance time $t^{\le}(z)$ is the largest time, for which the trajectory is in the green upper-closed set, i.e., $t^{\le}(z)=4$. Similarly, the robust upper dominance time $t^{\ge}(z)$ is the largest time, for which the trajectory is in the red lower-closed set, i.e., $t^{\le}(z)=\infty$. }
        \label{fig:hittingtime}
        \vspace{-0.5cm}
        \end{figure}

The robust upper dominance time $t^{\le}$ is the largest time $t \ge 0$ such that $\testx(t)$ satisfies the order relation $x \le \testx(t) + \lipconst{t-1}\vect{1}_n$, or equivalently, the largest time $t \ge 0$ such that the inflated trajectory $\testx(t) + \lipconst{t-1}\vect{1}_n$ belongs to the robust upper-closed set $[\{x\}]_{\uparrow}$. This notion is reminiscent of the \emph{hitting-time function} in the literature on stochastic processes~\cite[Chapter 10]{DAL-YP:17}. However, unlike the hitting time---which characterizes the \emph{first} occurrence of an event---condition~\eqref{eq:Phattime} instead captures the \emph{last} time at which the order relation $x \le \testx(t) + \lipconst{t-1}\vect{1}_n$ holds (we refer to Fig. \ref{fig:hittingtime} for an illustration of robust upper dominance times and functions associated with a trajectory of a $2$-dimensional linear system).

\begin{definition}[Robust Lower Dominance Functions]\label{def:adisshatfcn}
Let $\mS=(\mX, \mX_0,\mW, f)$ be an SM system satisfying Assumption~\ref{as:lipschitz}. Given a trajectory $\testx$ generated under an arbitrary disturbance and initial condition $\testx(0) \in \mX$, the \emph{robust lower dominance time} $t^{\geq} : \mX \to \Re_{\geq 0} \cup \{\infty\}$ is defined by
\begin{align}
t^{\geq}(x) &:= \sup \left\{t\mid x\geq \testx(t) - \lipconst{t-1}\vect{1}_n\right\}.\label{eq:Qhattime}
\end{align}
where $\lipconst{t-1}$ is as defined in~\eqref{eq:compact}. Then, for a given parameter $\alpha\in \Re_{>1}$, the \textit{robust lower dominance function} $\adiss^{\testx}:\mathcal{X}\to \Re$ is defined by
\begin{align}
\adiss^{\testx} (x) &:= \begin{cases}
     \tfrac{1}{t^{\geq}(x)+1} & t^{\geq}(x) < \infty,\\
     0& t^{\geq}(x)= \infty,\\
     \alpha & t^{\geq}(x)= \emptyset.
 \end{cases}\label{eq:adisshatfcn}
\end{align}    
\end{definition}
\smallskip

The robust lower dominance time $t^{\ge}$ is the largest time $t \ge 0$ such that $\testx(t)$ satisfies the order relation $x \ge \testx(t) + \lipconst{t-1}\vect{1}_n$ or, equivalently, the largest time $t \ge 0$ such that $\testx(t)+\lipconst{t-1}\vect{1}_n$ belongs to the lower-closed set $[\{x\}]_{\downarrow}$ (see Fig. \ref{fig:hittingtime} for an illustration of robust lower dominance times and functions associated with a trajectory of a $2$-dimensional linear system).
\begin{remark}[Robust Dominance Functions]
\;
\begin{enumerate}
    \item (\textit{Robustness margin}) The definitions of the robust dominance functions $\diss^{\testx}$ and $\adiss^{\testx}$ are based on an \emph{inflated} trajectory of the form $\testx(t) \pm \lipconst{t-1}\vect{1}_n$. Indeed, by Proposition~\ref{thm:traj_comp}, this inflation acts as a robustness margin with respect to disturbances in our safety verification procedure (see Theorem~\ref{thm:robustsafety}).
    \item (\textit{Data-driven computations}) By Definitions~\ref{def:disshatfcn} and~\ref{def:adisshatfcn}, the robust dominance functions $\diss^{\testx}$ and $\adiss^{\testx}$ can be evaluated at any $x \in \mX$ using only the collected trajectory $\testx$, the Lipschitz bounds $L_x$, $L_w$, and the disturbance bound $D_w$, without requiring knowledge of the state-transition map $f$. This property makes these functions particularly well-suited for data-driven robust safety verification.
\end{enumerate} 
\end{remark}
\smallskip 

By exploiting the monotonicity of $\mS$, the next theorem establishes key properties of the robust dominance functions.

\begin{theorem}[Properties of Dominance functions]\label{thm:a-disshatfcn}
Consider an SM system $\mS=(\mX,\mX_0,\mW,f)$ that satisfies Assumptions~ \ref{as:test_verification} and \ref{as:lipschitz}, and let $\testx$ be a trajectory of the system. Then, the following statements hold:
\begin{enumerate}
   \item\label{ph1}\textit{(Monotonicity)} for every $x,y \in \mX$, we have 
   \begin{align*}
     x\leq y \quad\implies \quad \diss^{\testx}(x) \leq \diss^{\testx}(y),\quad \adiss^{\testx}(x) \geq \adiss^{\testx}(y).
   \end{align*}
   i.e., $\diss^{\testx}$ and $-\adiss^{\testx}$ are monotone on $\mX$.
   \item\label{ph2} \textit{(Dissipation)} for every $x\in \mX$,
    \begin{align*}
      \diss^{\testx}(f(x,w)) &\le \diss^{\testx}(x), &&\mbox{ for all } w \in \mW,\\
      \adiss^{\testx}(f(x,w)) &\le \adiss^{\testx}(x), &&\mbox { for all } w \in \mW.
    \end{align*}
   \item\label{ph3}\textit{(Invariance)} for every $c\in [0,\alpha]$, the $c$-sublevel sets 
    \begin{align*}
     \left(\diss^{\testx}\right)_{\le c}&:=\{x \in \mX \mid \diss^{\testx}(x) \leq c \},\\
     \left(\adiss^{\testx}\right)_{\le c}&:=\{x \in \mX \mid \adiss^{\testx}(x) \leq c \}
    \end{align*}
are robust invariant sets for $\mS$.
\end{enumerate}
\end{theorem}
\smallskip
\paragraph*{Proof: }See Appendix \ref{ap:thmadisshatfcn}.
$\hfill\blacksquare$
\smallskip
\begin{remark}[Barriers via Dominance functions]\;
\begin{enumerate}
    \item \textit{(Dissipation property and RBCs)}. Theorem~\ref{thm:a-disshatfcn} part~\ref{ph2} shows that the robust dominance functions decrease inductively along the system’s evolution under any disturbances. This dissipation property resembles condition~\eqref{eq:rbc3} for RBCs and motivates the use of these functions as building blocks for constructing RBCs for $\mS$, as will be shown in Subsection~\ref{sec:TBRBC}.
    \item \textit{(Upper- and lower-closed invariant sets)}. The invariance property of the robust dominance functions follows directly from their dissipation property. Indeed, due to the monotonicity of $\diss^{\testx}$ and $-\adiss^{\testx}$, the sets $\left(\diss^{\testx}\right)_{\le c}$ and $\left(\adiss^{\testx}\right)_{\le c}$ are lower-closed and upper-closed, respectively. 
\end{enumerate}
\end{remark}
\subsection{Trajectory-based Robust Barrier Certificates}\label{sec:TBRBC}

The resemblance between the dissipation property of the robust dominance functions (see Theorem~\ref{thm:a-disshatfcn}, part~\eqref{ph2}) and condition~\eqref{eq:rbc3} for RBCs suggests that these functions are natural candidates for constructing RBCs for state-monotone systems. The next theorem shows that, for systems without disturbances, robust safety with respect to a certain class of unsafe sets can be completely characterized by robust barrier functions constructed from robust dominance functions.
\begin{theorem}[RBCs via Robust dominance functions]\label{thm:RBC-mon}
Consider an SM system $\mS=(\mX,\mX_0,\mW,f)$ with $\mW=\emptyset$ and an unsafe set $\mX_u\subseteq \mX$. Suppose that the unsafe set $\mX_u$ can be expressed as a union of upper-closed and lower-closed sets. Let $\Sigma$ denote the set of all trajectories of $\mS$ starting from $\mX_0$. For a given trajectory $\hatx\in\Sigma$, let $\diss^{\hatx}$ and $\adiss^{\hatx}$ be the robust upper and lower dominance functions associated with $\hatx$, as introduced in Definitions~\ref{def:disshatfcn} and~\ref{def:adisshatfcn}, respectively. Then, the following statements are equivalent:
\begin{enumerate}
\item\label{thm:RBC-mon1} the system $\mS$ is robustly safe with respect to $\mX_u$,
\item\label{thm:RBC-mon2} the function $\mathbb{B}:\mX\to \Re$ defined by
\begin{align}\label{eq:RBC-mon}
\barrier(x) = \inf_{\hatx\in \Sigma} \left\{\max\left\{\diss^{\hatx}(x) , \adiss^{\hatx}(x)\right\}\right\}-1
\end{align}
is an RBC for the system $\mS$ with respect to $\mX_u$.
\end{enumerate}
\end{theorem}
\smallskip
\paragraph*{Proof: }See Appendix \ref{ap:thm:RBC-mon}. 
$\hfill\blacksquare$
\smallskip

For SM discrete-time systems without disturbances and with unsafe sets which is union of lower and upper closed sets, Theorem~\ref{thm:RBC-mon} shows that robust upper and lower dominance functions are sufficiently expressive to construct RBC and can, therefore, be viewed as \emph{basis functions} for RBC construction.
Despite this complete characterization, Theorem~\ref{thm:RBC-mon} is an existential and non-constructive result. Indeed, the robust barrier function in~\eqref{eq:RBC-mon} is defined in terms of \textit{infinitely many} robust upper and lower dominance functions, whereas in practice only finitely many system trajectories are available.

For SM discrete-time systems with disturbances with arbitrary unsafe sets, inspired by Theorem~\ref{thm:RBC-mon}, we seek a tractable robust barrier certificate expressed as an affine combination of robust dominance functions generated from a finite set of trajectories. Given a finite set of collected trajectories $\testx_k$, $k \in [1;N]$, we use the corresponding functions $\diss^{\testx_k}$ and $\adiss^{\testx_k}$ to build a candidate robust barrier certificate $\barrier^{\testx_{[1;N]}}:\mX \to \Re$ for the state monotone system $\mS$ as follows:
\begin{align}\label{eq:Tbarrier}
\barrier^{\testx_{[1;N]}}(x) := a + \sum_{k=1}^N \left(b_k \diss^{\testx_k}(x) + c_k\adiss^{\testx_k}(x)\right).
\end{align}
We call $\barrier^{\testx_{[1;N]}}$ a \textit{Trajectory-based Robust Barrier Certificate (T-RBC)} if it satisfies:
\begin{subequations}\label{eq:trbc}
\begin{align}
   \barrier^{\testx_{[1;N]}}(x) &\leq 0, &&\mbox{ for all } x \in \mX_0,\label{eq:trbc1}\\
   \barrier^{\testx_{[1;N]}}(x) &> 0, &&\mbox{ for all } x \in \mX_u,\label{eq:trbc2}\\
   b_k,c_k &\geq 0, &&\mbox{for all } k \in [1;N].\label{eq:trbc3}
\end{align}
\end{subequations}

We next show that T-RBCs constructed from a finite collection of collected trajectories can be used to certify the robust safety of an unknown state monotone system.

\begin{theorem}[T-RCBs for robust safety]\label{thm:robustsafety}
Consider an SM system $\mS = (\mX, \mX_0, \mW, f)$ with an unsafe set $\mX_u \subseteq \mX$. Suppose the state-transition map $f$ is unknown but satisfies Assumption~\ref{as:lipschitz}, and that a finite collection of trajectories ${\testx_1,\ldots,\testx_N}$ has been collected satisfying Assumption~\ref{as:test_verification}. If there exists a T-RBC $\mathbb{B}^{{\testx}_{[1;N]}}$ of the form~\eqref{eq:Tbarrier} that satisfies conditions~\eqref{eq:trbc}, then
\begin{enumerate}
    \item $\mathbb{B}^{{\testx}_{[1;N]}}_{\le 0} = \left\{x\in \mX\mid \mathbb{B}^{{\testx}_{[1;N]}}(x)\le 0\right\}$ is a robust forward invariant set for $\mS$, and 
    \item system $\mS$ is robustly safe with respect to $\mX_u$. 
\end{enumerate}
\end{theorem}
\smallskip
\paragraph*{Proof: }
We prove that if we can find a function $\mathbb{B}^{{\testx}_{[1;N]}}$ of the form~\eqref{eq:Tbarrier} satisfying conditions~\ref{eq:trbc}, then $\mathbb{B}^{{\testx}_{[1;N]}}$ is an RBC, as in Definition \ref{def:robust_barrier}, for the system $\mS$. The result then follows, since the existence of an RBC is sufficient to certify the robust safety of the system~\cite{Prajna2004}. It is easy to show that equations \eqref{eq:trbc1} and \eqref{eq:trbc2} are equivalent to \eqref{eq:rbc1} and \eqref{eq:rbc2}, respectively. By Theorem \ref{thm:a-disshatfcn} part~\ref{ph2}, the robust upper and lower dominance function $\diss^{\testx_k}$ and $\adiss^{\testx_k}$ are dissipative, for every $k\in [1;N]$. As a result, for all $w \in \mW$, we have
\begin{align*}
  &a + \sum_{k=1}^N \left(b_k \diss^{\testx_k}(f(x,w)) + c_k\adiss^{\testx_k}(f(x,w))\right) 
  \leq\\
  &\quad \qquad a + \sum_{k=1}^N \left(b_k \diss^{\testx_k}(x) + c_k\adiss^{\testx_k}(x)\right),
\end{align*}
where in the above inequality we used the fact that $b_k,c_k\ge 0$ by condition \eqref{eq:trbc3}. 

This implies that $\barrier^{\testx_{[1;N]}}(f(x,w))\leq \barrier^{\testx_{[1;N]}}(x)$ for all $w \in \mW$, which is condition \eqref{eq:rbc3}.
$\hfill\blacksquare$
\smallskip
\begin{remark}[Trajectory-based RBCs]\;
\begin{enumerate}
    \item (\textit{Data-driven Barriers}) Theorem~\ref{thm:robustsafety} provides a systematic framework for verifying the robust safety of an unknown monotone system by combining the basis functions $\diss^{\testx}$ and $\adiss^{\testx}$ associated to collected trajectories. In contrast to the RBC conditions in~\eqref{eq:rbc}, the trajectory-based barrier certificates in~\eqref{eq:trbc} require no knowledge of the state-transition map and are constructed entirely from collected trajectories. This shift from a model-based condition to a data-driven one is enabled by the dissipation properties of the robust upper and lower dominance functions, which convert the model-based RBC condition~\eqref{eq:rbc3} into the model-free constraint~\eqref{eq:trbc3} on the coefficients $b_k,c_k$.
    
    \item (\textit{Computation of T-RBCs}) Verifying robust safety of an SM system using Theorem~\ref{thm:robustsafety} requires solving the T-RBC conditions~\eqref{eq:trbc1}--\eqref{eq:trbc3} for the coefficients $a,{b_k},{c_k}$. This task is challenging because closed-form expressions for the dominance functions $\diss^{\testx_k}$ and $\adiss^{\testx_k}$ are typically difficult to obtain, rendering many standard approaches—such as SOS programming and SMT-based methods—inapplicable. In Section~\ref{sec:datadriven}, we address this challenge by developing a sampling-based approach that leverages the monotonicity of the dominance functions.
\end{enumerate}
\end{remark}
\section{Trajectory-based Control Synthesis}\label{sec:control}

In this section, we focus on Problem~\ref{prob2} and establish a framework using collected trajectories of an unknown SIM system $\mS$ to certify controlled safety and synthesize a controller that guarantees safety.
Our central idea is to construct trajectory-based barrier certificates analogous to those introduced in Section~\ref{sec:verification}. The key distinction is that each basis element is paired with an associated control input to capture the influence of control actions along the corresponding trajectory.

\subsection{Controlled Dominance Functions}

In this section, we introduce two classes of data-driven functions associated with each collected trajectory, which serve as building blocks for our controlled safety certificates.
\begin{definition}[Controlled Upper Dominance functions]\label{def:dissfcn}
Consider an SIM system $\mS=(\mX, \mX_0,\mU, f)$ and let $\testx$ be a trajectory of $\mS$ given monotone controller $\ctrl:\mX \to \mU$ and initial condition $\testx(0) \in \mX$, the \emph{controlled upper dominance time} is $t^{\leq}_{\ctrl} : \mX \to \Re_{\ge 0} \cup \{\infty\}$ defined by
\begin{align}
t_{\ctrl}^{\leq}(x)&:=\sup \left\{t\mid x \leq \testx(t) \right\}.\label{eq:Ptime}
\end{align}
Then, for a given parameter $\alpha \in \Re_{>1}$, the \textit{controlled upper dominance function} $\diss_{\ctrl}^{\testx}:\mX \to \Re_{\ge 0}$ is defined by
\begin{align}\label{eq:dissfcn}
 \diss_{\ctrl}^{\testx} (x)&:= \begin{cases}
     \tfrac{1}{t_{\ctrl}^{\leq}(x)+1} & t_{\ctrl}^{\leq}(x) < \infty,\\
     0 & t_{\ctrl}^{\leq}(x) = \infty,\\
     \alpha & t_{\ctrl}^{\leq}(x) = \emptyset.
 \end{cases}
\end{align}
\end{definition}
\smallskip
\begin{definition}[Controlled Lower Dominance functions]\label{def:adissfcn}
Consider an SIM system $\mS=(\mX, \mX_0,\mU, f)$ and let $\testx$ be a trajectory of $\mS$ given monotone controller $\ctrl:\mX \to \mU$ and initial condition $\testx(0) \in \mX$, the \emph{controlled lower dominance time} is $t_{\ctrl}^{\geq} : \mX \to \Re_{\ge 0} \cup \{\infty\}$ defined by
\begin{align}
t_{\ctrl}^{\geq}(x) &:= \sup \left\{t\mid x\geq \testx(t)\right\}.\label{eq:Qt}
\end{align}
Then, for a given parameter $\alpha \in \Re_{>1}$, the \textit{controlled lower dominance function} $\adiss_{\ctrl}^{\testx}:\mX \to \Re_{\ge 0}$ is defined by
\begin{align}
\adiss_{\ctrl}^{\testx} (x) &:= \begin{cases}
     \tfrac{1}{t_{\ctrl}^{\geq}(x)+1} & t_{\ctrl}^{\geq}(x) < \infty,\\
     0 & t_{\ctrl}^{\geq}(x)= \infty,\\
     \alpha & t_{\ctrl}^{\geq}(x)= \emptyset.
 \end{cases}\label{eq:adissfcn}
\end{align}    
\end{definition}
The definitions of the controlled dominance functions are similar to that of the robust dominance functions in~\eqref{eq:disshatfcn} and~\eqref{eq:adisshatfcn} in the case $\mW = \emptyset$. The key distinction between the two lies in how the system input is interpreted.
In the robust dominance functions, the trajectory $\testx$ is interpreted as being generated under an \emph{unknown} disturbance. In contrast, in the controlled dominance functions, the trajectory $\testx$ is regarded as a controlled trajectory driven by a \emph{known} controller $\ctrl$.

By exploiting the monotonicity of the state-transition map, in the next theorem we investigate the fundamental properties of the controlled upper and lower dominance functions.

\begin{theorem}[Properties of Controlled Dominance Functions]\label{thm:a-dissfcn} Consider an SIM system $\mS=(\mX,\mX_0,\mU,f)$ that satisfies Assumption \ref{as:test_control} and let $\testx$ be a trajectory of $\mS$ given by monotone feedback controller $\ctrl:\mathcal{X}\to \mathcal{U}$. Then, the following statements hold:
\begin{enumerate}
   \item\label{p1}\textit{(Monotonicity)} for every $x,y \in \mX$, we have 
   \begin{align*}
     x\leq y \quad\implies \quad \diss_{\ctrl}^{\testx}(x) \leq \diss_{\ctrl}^{\testx}(y),\;\;\;\; \adiss_{\ctrl}^{\testx}(x) \geq \adiss_{\ctrl}^{\testx}(y),
   \end{align*}
   i.e., $\diss_{\ctrl}^{\testx}$ and $-\adiss_{\ctrl}^{\testx}$ are monotone on $\mX$.
   \item\label{p2} \textit{(Dissipation)} for every $x\in \mX$, we have 
    \begin{align*}
      &\diss_{\ctrl}^{\testx}(f(x,u)) \le \diss_{\ctrl}^{\testx}(x), &&\qquad\mbox{for all } u \in  [\{\ctrl(x)\}]_{\downarrow},\\
      &\adiss_{\ctrl}^{\testx}(f(x,u)) \le \adiss_{\ctrl}^{\testx}(x), &&\qquad \mbox{for all }u \in [\{\ctrl(x)\}]_{\uparrow}.
    \end{align*}
   \item\label{p3}\textit{(Invariance)} for every $c\in [0,\alpha]$, the $c$-sublevel sets 
    \begin{align*}
     \left(\diss^{\testx}_{\ctrl}\right)_{\le c}:=\{x \in \mX \mid \diss_{\ctrl}^{\testx}(x) \leq c \},\\
     \left(\adiss^{\testx}_{\ctrl}\right)_{\le c}:=\{x \in \mX \mid \adiss_{\ctrl}^{\testx}(x) \leq c \}
    \end{align*}
are controlled invariant set for $\mS$.
\end{enumerate}
\end{theorem}
\paragraph*{Proof: } See Appendix \ref{ap:thmadissfcn}.
$\hfill\blacksquare$

\subsection{Trajectory-based Control Barrier Certificates}
The resemblance between the dissipation property of the controlled dominance functions (see Theorem~\ref{thm:a-dissfcn}, part~\eqref{p2}) and condition~\eqref{eq:cbc3} for CBCs suggests that these functions are natural candidates for constructing CBCs for input- and state-monotone systems.
Motivated by this observation, in this section we develop a data-driven framework for studying controlled safety of SIM systems using control barrier certificates expressed as affine combinations of robust dominance functions generated from a finite collection of trajectories.
Given a finite family of collected trajectories $\testx_k$ and their corresponding controllers $\ctrl_k$, we use the associated controlled dominance functionals $\diss_{\ctrl_k}^{\testx_k}$ and $\adiss_{\ctrl_k}^{\testx_k}$, defined in~\eqref{eq:dissfcn} and~\eqref{eq:adissfcn}, as basis functions. Based on these functionals, we construct the following CBC candidate:
\begin{align}
 \barrier^{\testx_{[1;N]}}(x) &:= a + \sum_{k=1}^N \left(b_k \diss_{\ctrl_k}^{\testx_k}(x) + c_k\adiss_{\ctrl_k}^{\testx_k}(x)\right),\label{eq:Tcbarrier1}
 \end{align}
with the safe feedback controller $\pi:\mX\to \mU$ given by
\begin{align} 
\pi(x) \in \Pi(x)&:= \bigcap_{k^p \in K^p}[\{\ctrl_{k^p}(x)\}]_{\downarrow}\bigcap_{k^q \in K^q}[\{\ctrl_{k^q}(x)\}]_{\uparrow},\label{eq:Tcbarrier2} 
\end{align}
with $K^p := \{k \in [1;N] \mid b_k > 0\}$ and $K^q := \{k \in [1;N] \mid c_k > 0\}$. We say that $\barrier^{\testx_{[1;N]}}$ defined in~\eqref{eq:Tcbarrier1} with controller satisfying~\eqref{eq:Tcbarrier2} is a \textit{Trajectory-based Control Barrier Certificate (T-CBC)} if:
\begin{subequations}\label{eq:tcbc}
\begin{align}
   \barrier^{\testx_{[1;N]}}(x) &\leq 0, &&\mbox{for all } x \in \mX_0,\label{eq:tcbc1}\\
   \barrier^{\testx_{[1;N]}}(x) &> 0, &&\mbox{for all } x \in \mX_u,\label{eq:tcbc2}\\
   b_k,c_k &\geq 0, &&\mbox{for all } k \in [1;N];\label{eq:tcbc3}
\end{align}
\end{subequations}
and, for all $ x \in \barrier^{\testx_{[1;N]}}_{\leq 0}$, we have that
\begin{align}\label{eq:tcbc_u}
    \Pi(x) \neq \emptyset. \tag{21d}
\end{align}

The next result shows that T-CBCs constructed from collected trajectories with monotone feedback policies can be used to safely control an SM monotone system.

\begin{theorem}[T-CBCs provide safe controllers]\label{thm:controlsafety}
Consider an SIM system $\mS=(\mX, \mX_0,\mU, f)$ with an unknown state-transition map $f$ together with a finite collection of trajectories ${\testx_1,\ldots,\testx_N}$ given by monotone feedback controllers $\ctrl_1,\ldots,\ctrl_N$ satisfying Assumption~\ref{as:test_control}, and a given unsafe set $\mX_u \subseteq \mX$. If there exists a T-CBC $\barrier^{\testx_{[1;N]}}$ of the form~\eqref{eq:Tcbarrier1} that satisfies conditions~\eqref{eq:tcbc}, then for any feedback controller $\pi:\mX\to \mU$ satisfying~\eqref{eq:Tcbarrier2},
\begin{enumerate}
    \item the set $\barrier^{\testx_{[1;N]}}_{\leq 0}=\{x\in \mX\mid \barrier^{\testx_{[1;N]}}(x) \le 0\}$ is controlled invariance for $\mS$, and 
    \item system $\mS$ is controlled safe with respect to $\mX_u$. 
\end{enumerate}
\end{theorem}
\smallskip
\paragraph*{Proof: } 
We prove that a T-CBC of the form $\barrier^{\testx_{[1;N]}}$ satisfying conditions~\eqref{eq:tcbc} is a CBC for the system $\mS$ as in Definition \ref{def:cbarrier}. Notice that equations \eqref{eq:tcbc1} and \eqref{eq:tcbc2} are equivalent to \eqref{eq:cbc1} and \eqref{eq:cbc2}, respectively. By condition \eqref{eq:tcbc3}, function $\barrier^{\testx_{[1;N]}}$ in \eqref{eq:Tcbarrier1} preserves the properties of each $\diss_{\ctrl_k}^{\testx_k}$ and $\adiss_{\ctrl_k}^{\testx_k}$ stated in Theorem \ref{thm:a-dissfcn}, in particular, \textit{dissipation}. Then, by condition \eqref{eq:tcbc_u}, the map \eqref{eq:Tcbarrier2} provides a set of control inputs that guarantees the existence of $u = \pi(x) \in \Pi(x)$, such that $\barrier^{\testx_{[1;N]}}(f(x,u))\leq \barrier^{\testx_{[1;N]}}(x)$.
$\hfill\blacksquare$
\smallskip
\begin{remark}[Trajectory-based CBCs]\;
\begin{enumerate}
    \item (\textit{Choice of Safe Controller}) The function $\Pi$ in~\eqref{eq:Tcbarrier2} is set-valued, and any feedback controller $\pi$ that is a single-valued selection from $\Pi(x)$ guarantees the safety of the system $\mS$. This flexibility in choice of control policy is particularly advantageous when the safe controller is deployed as a safety shield that modifies a nominal controller: one can select, at each state $x$, a control input from $\Pi(x)$ that remains as close as possible to the nominal input, thereby ensuring safety while minimally altering the original control objective.
    
    \item (\textit{Choice of Collected Trajectories}) The selection and number of collected trajectories directly influence both the expressiveness of the CBCs and the admissible range of control actions. Increasing the number of trajectories (larger $N$) provides a richer collection of basis functions $\diss_{\ctrl_k}^{\testx_k}$ and $\adiss_{\ctrl_k}^{\testx_k}$. On the other hand, it also increases the number of nonzero coefficients $b_k,c_k$ in the representation~\eqref{eq:Tcbarrier1}, making it significantly more challenging to satisfy condition~\eqref{eq:tcbc_u}. As a result, the admissible control set may shrink substantially or even become empty.
\end{enumerate}

\end{remark}

\section{Sampling-based Barrier Certificates}\label{sec:datadriven}

Although Theorems~\ref{thm:robustsafety} and~\ref{thm:controlsafety} provide an elegant trajectory-based framework for addressing Problems~\ref{prob1} and~\ref{prob2} through T-RBCs and T-CBCs, they do not offer a systematic procedure for computing these certificates from the collected trajectories.
In this section, we develop an efficient sampling-based approach for computing T-RBCs and T-CBCs.

We begin by formulating the search for T-RBCs (resp. T-CBCs) as an optimization problem. Given a template for T-RBC (resp. T-CBCs) $\barrier^{\testx_{[1;N]}}$ as in~\eqref{eq:Tbarrier} (resp.~\eqref{eq:Tcbarrier1}), we collect the coefficients $a,b_k,c_k$, for $k \in [1;N]$, into a parameter vector $p := (a,b_1,\ldots,b_N,c_1,\ldots,c_N)\in \mathcal{P}$, where $\mc{P}$ denotes the admissible parameter set. We then consider the following optimization problem to search for T-RBCs (resp. T-CBCs) of the monotone system $\mS$:
\begin{equation}\label{eq:OP}\tag{\textbf{OP}}
\begin{aligned}\min_{p \in \mc{P}} \quad &\mathsf{Loss}(p)\\
&\mbox{subject to }\; \eqref{eq:trbc}\;\; (\mbox{resp.}~\eqref{eq:tcbc}), 
\end{aligned}    
\end{equation}
where $\mathsf{Loss}$ is a prescribed loss function—e.g., $0$, $\|p\|_1$, or any cost designed to promote desirable properties of $p$, such as sparsity or reduced control effort.

Despite its simplicity, the optimization problem~\eqref{eq:OP} is difficult to solve in practice for two reasons. 
First, the T-CBC template $\barrier^{\testx_{[1;N]}}$ depends on dominance functions, which are defined in terms of infinite-length trajectories $\testx$. In practice, however, only finite trajectory segments can be collected.
Second, the dominance functions generally do not admit closed-form expressions. Consequently, symbolic constraint-satisfaction approaches—such as Sum-of-Squares (SOS) methods~\cite{Prajna2004} and SMT-based methods~\cite{edwards2024fossil}—cannot be directly used to enforce conditions~\eqref{eq:trbc} (resp. conditions~\eqref{eq:tcbc}).

Motivated by the limitations discussed above, we develop a \textit{sampling-based} approach for solving the optimization problem~\eqref{eq:OP}.
The method consists of three main steps. First, we introduce the notion of truncated dominance functions constructed from \textit{finite-length trajectories}. We then impose two different technical assumptions on the collected trajectories and study the properties of the truncated dominance functions under each setting. Although these assumptions are incomparable, they share the same objective: \textit{to dominate the unobserved tails of trajectories that cannot be collected}.
Finally, by exploiting the monotonicity of the dominance functions, we develop a method for enforcing conditions~\eqref{eq:trbc} (resp.~\eqref{eq:tcbc}) using only finitely many sampled states.

\subsection{Truncated Robust Dominance Functions}

In this section, we introduce truncated robust dominance functions, a finite-horizon variant of robust dominance functions that can be computed from finite-length trajectories. 
We consider the following assumption.
\begin{assumption}[Compact Tail]\label{as:convergence}
For every collected trajectory $\testx$ of the unknown SM system $\mS=(\mX,\mX_0,\mW,f)$, there exists $\eps_T \in \Re_{>0}$ such that
\begin{align}
    &\testx(T) - \eps_T \vect{1}_n \leq \testx(t) \leq \testx(T) + \eps_T\vect{1}_n,\label{eq:epsilonT}
\end{align}
for every $t\in \NN_{\geq T}$.
\end{assumption}

Note that Assumption \ref{as:convergence} is fulfilled if each collected trajectory converges to a point in $\mX$. This condition allows us to address the loss of information induced by using the $T$-truncated trajectories in upper and lower dominance functions.
To show this effect, we introduce the $T$-truncated upper dominance functions $\diss^{\testx^T}:\mX\to \Re_{\ge 0}$ as follows:
\begin{subequations}\label{eq:rvdissT}
\begin{align}
\diss^{\testx^T}(x) &:= \begin{cases}
    \tfrac{1}{t^{\leq_T}(x) + 1} & t^{\leq_T}(x) \leq T,\\
    \alpha, & t^{\leq_T}(x) =\emptyset,
\end{cases}\\ 
t^{\leq_T}(x)&:=\max_{t \in [0;T]} \left\{t\mid x -\eps_T\vect{1}_n\leq \testx(t)+ \lipconst{t-1}\vect{1}_n \right\},
\end{align}     
\end{subequations}
where $\eps_T$ is given in~\eqref{eq:epsilonT}. Similarly, we introduce the $T$-truncated lower dominance functions $\adiss^{\testx^T}:\mX \to \Re_{\geq 0}$ as:
\begin{subequations}\label{eq:rvadissT}
\begin{align}
\adiss^{\testx^T}(x) &:= \begin{cases}
    \tfrac{1}{t^{\geq_T}(x) + 1} & t^{\geq_T}(x) \leq T,\\
    \alpha & t^{\geq_T}(x) =\emptyset.
\end{cases}\\
t^{\geq_T}(x)&:=\max_{t \in [0;T]} \left\{t\mid x +\eps_T\vect{1}_n\geq \testx(t)- \lipconst{t-1}\vect{1}_n\right\},
\end{align}     
\end{subequations}
where $\eps_T$ is given in~\eqref{eq:epsilonT}.
The $T$-truncated upper and lower dominance functions can be computed entirely from finite-length collected trajectories $\testx^T \in \mX^{T}$. We next show that these truncated functions are monotone, and establish upper and lower bounds on the robust dominance functions in terms of their $T$-truncated counterparts.

\begin{theorem}[Truncated Robust Dominance Functions]\label{thm:rvPQT}
Consider an SM system $\mS=(\mX,\mX_0,\mW,f)$ that satisfies Assumptions~\ref{as:lipschitz} and \ref{as:convergence}, and let $\testx$ be a trajectory of the system. Then, for every $T\ge 0$, the following statements hold:
 \begin{enumerate}
     \item\label{p1-trunc-rv} $\diss^{\testx^T}$ and $-\adiss^{\testx^T}$ are monotone functions, 
     \item\label{p2-trunc-rv} for every $x\in \mX$,
 \begin{align*}
     \diss^{\testx^T}(x) - \tfrac{1}{T+1}\leq \diss^{\testx}(x) \leq \diss^{\testx^T}(x + \eps_T \vect{1}_n),
 \end{align*}
 \begin{align*}
     \adiss^{\testx^T}(x) - \tfrac{1}{T+1} \leq \adiss^{\testx}(x) \leq \adiss^{\testx^T}(x + \eps_T \vect{1}_n), 
 \end{align*}
 where $\eps_T$ is as defined in~\eqref{eq:epsilonT}.
 \end{enumerate}
\end{theorem}
\smallskip
\paragraph*{Proof: }See Appendix \ref{ap:rvthmPQT}.
$\hfill\blacksquare$

\subsection{Truncated Controlled Dominance Functions}

In this section, we introduce a finite-horizon variant of the controlled dominance functions, called truncated controlled dominance functions, which can be computed from finite-length trajectories. We consider the following assumption on the tail of the collected trajectories.
\begin{assumption}[Dominating tail]\label{as:finalxT}
For each collected trajectory $\testx$ of the system $\mS=(\mX,\mX_0,\mU,f)$ in Assumption~\ref{as:test_control}, there exists a finite time horizon $T>0$ such that one of the following conditions hold:
\begin{enumerate}
    \item \label{as1:finalxTP}$\testx(T) \leq \testx(T-1)$, or
    \item \label{as1:finalxTQ}$\testx(T) \geq \testx(T-1)$.
\end{enumerate}
\end{assumption}
We now define the \textit{$T$-truncated upper dominance function} $\diss_{\ctrl}^{\testx^T}:\mX\to \Re_{\ge 0}$ as follows:
\begin{subequations}\label{eq:dissT}
\begin{align}
t_{\ctrl}^{\leq_T}(x)&:=\max_{t \in [0;T-1]} \left\{t\mid x \leq \testx^T(t)\right\}\\
\diss_{\ctrl}^{\testx^T}(x) &:= \begin{cases}
    \tfrac{1}{t_{\ctrl}^{\leq_T}(x) + 1} & t_{\ctrl}^{\leq_T}(x) \leq T,\\
    \alpha, & t_{\ctrl}^{\leq_T}(x) =\emptyset,
\end{cases}
\end{align}     
\end{subequations}
where $\alpha\in \Re_{> 1}$. Similarly, we introduce the \textit{$T$-truncated lower dominance function} $\adiss_{\ctrl}^{\testx^T}:\mX \to \Re_{\geq 0}$ as:
\begin{subequations}\label{eq:adissT}
\begin{align}
t_{\ctrl}^{\geq_T}(x)&:=\max_{t \in [0;T-1]} \left\{t\mid x \geq \testx(t)\right\}\\
\adiss_{\ctrl}^{\testx^T}(x) &:= \begin{cases}
    \tfrac{1}{t_{\ctrl}^{\geq_T}(x) + 1} & t_{\ctrl}^{\geq_T}(x) \leq T,\\
    \alpha & t_{\ctrl}^{\geq_T}(x) =\emptyset.
\end{cases}
\end{align}     
\end{subequations}

Using Assumption~\ref{as:finalxT}, we establish monotonicity, dissipation, and invariance properties for the truncated dominance functions $\diss_{\ctrl}^{\testx^T}$ and $\adiss_{\ctrl}^{\testx^T}$, analogous to Theorem~\ref{thm:a-dissfcn}.

\begin{theorem}[Truncated Controlled Dominance Functions]\label{thm:PQT}
Consider an SIM system $\mS=(\mX,\mX_0,\mU,f)$ and let $\testx$ be a trajectory of $\mS$ given by a monotone feedback controller $\ctrl:\mathcal{X}\to \mathcal{U}$. F
or every $T\ge 0$,
 \begin{enumerate}
     \item\label{p1-trunc-cs} If Assumption \ref{as:finalxT}, part \ref{as1:finalxTP} holds, then: $\diss_{\ctrl}^{\testx^T}$ is a monotone function; for every $x \in \mX$, we have
     \begin{align*}
     \diss_{\ctrl}^{\testx^T}(f(x,u)) \leq \diss_{\ctrl}^{\testx^T}(x), \qquad \mbox{for all } u \in [\{\ctrl(x)\}]_{\downarrow};    
     \end{align*}
     and for every $c\in [0,\alpha]$, the $c$-sublevel set 
     $$\left(\diss_{\ctrl}^{\testx^T}\right)_{\le c}:=\{x \in \mX \mid \diss_{\ctrl}^{\testx^T}(x) \leq c \}$$ 
     is a controlled invariant set for $\mS$.
     \item\label{p2-trunc-cs} If Assumption \ref{as:finalxT}, part \ref{as1:finalxTQ} holds, then: $-\adiss_{\ctrl}^{\testx^T}$ is a monotone function; for every $x \in \mX$, we have
      \begin{align*}
     \adiss_{\ctrl}^{\testx^T}(f(x,u)) \leq \adiss_{\ctrl}^{\testx^T}(x), \qquad \mbox{for all } u \in [\{\ctrl(x)\}]_{\uparrow};
    \end{align*}
    and for every $c\in [0,\alpha]$, the $c$-sublevel set 
    \begin{align*}
     \left(\adiss_{\ctrl}^{\testx^T}\right)_{\le c}:=\{x \in \mX \mid \adiss_{\ctrl}^{\testx^T}(x) \leq c \}
    \end{align*}
    is a controlled invariant set for $\mS$.
 \end{enumerate}
\end{theorem}
\smallskip
\paragraph*{Proof: }See Appendix \ref{ap:thmPQT}.
$\hfill\blacksquare$

\subsection{Sampling-based Barrier Certificates}
We next leverage the monotonicity of truncated dominance functions to develop a sampling-based approach to solve the optimization problem~\eqref{eq:OP}.
 
We begin by introducing an assumption and the notion of hyper-rectangular partition.

\begin{assumption}[Compact State Set]\label{as:compact}
For an SM system $\mS=(\mX,\mX_0,\mW,f)$ or an SIM system $\mS'=(\mX,\mX_0,\mU,f)$, the state set $\mX$ is compact.
\end{assumption}

\begin{definition}[Hyper-rectangular partition]\label{def:partition}
Let $\mc{I}$ be a finite set of indices. Then, a family of hyper-rectangles $\{[\underline{x}^i,\overline{x}^i]\}_{i\in \mc{I}}$, is called a \emph{hyper-rectangular partition} of compact set $\mX$ if, for every $x \in \mX$, there exists $i\in \mc{I}$ such that $x \in [\underline{x}^i,\overline{x}^i]$ and for every $i,j\in \mc{I}$, such that $i\neq j$, we have $[\underline{x}^i,\overline{x}^i]\cap [\underline{x}^j,\overline{x}^j]=\emptyset$.
  \end{definition}

The key idea of our approach is to enforce the constraints~\eqref{eq:trbc} (resp.~\eqref{eq:tcbc}) in optimization problem~\eqref{eq:OP} at sample points taken from a hyper-rectangular partition, and then leverage the monotonicity of the dominance functions to guarantee that these constraints hold throughout the entire partition. Consider a hyper-rectangular partition $\{[\underline{x}^i,\overline{x}^i]\}_{i\in \mc{I}}$ on the state set $\mX$ of the SM system $\mS = (\mX,\mX_0,\mW,f)$ or the SIM system $\mS' = (\mX,\mX_0,\mU,f)$ satisfying Assumption \ref{as:compact} such that $\mc{I}_{0}$ and $\mc{I}_{u}$ are minimal subsets of $\mc{I}$ such that
     \begin{align}\label{eq:indices}
    \mX_0\subseteq\bigcup\limits_{i\in \mc{I}_{0}}[\underline{x}^i,\overline{x}^i],\qquad
    \mX_u\subseteq\bigcup\limits_{i\in \mc{I}_{u}}[\underline{x}^i,\overline{x}^i].
\end{align}
Consider the collected trajectories $\testx_1,\ldots,\testx_N$. For $k \in [1;N]$ satisfying assumption~\ref{as:test_verification}. We define the $T$-truncated trajectory barrier function $\barrier^{\testx^T_{[1;N]}}:\mathcal{P}\times\mX \to \Re$ analogous to $\barrier^{\testx_{[1;N]}}$ in~\eqref{eq:Tbarrier} or \eqref{eq:Tcbarrier1} as follows:
\begin{align}
 \barrier^{\testx^T_{[1;N]}}(p,x) &:= a + \sum_{k=1}^N \left(b_k \diss^{\testx^T_k}(x) + c_k\adiss^{\testx^T_k}(x)\right).\label{eq:Tbarrier3}
 \end{align}
 with $b_k,c_k \in \Re_{\ge 0}$ for all $k \in [1;N]$. Using the monotonicity of the $T$-truncated dominance functions $\diss^{\testx^T_k}$ and $-\adiss^{\testx^T_k}$ provided by either Theorem~\ref{thm:rvPQT} or Theorem~\ref{thm:PQT}, it follows that $\widehat{\barrier}^{\testx^T_{[1;N]}}(p,x,y)$ defined by
\begin{align}\label{eq:Tcbarrier1-data}
 \widehat{\barrier}^{\testx^T_{[1;N]}}(p,x,y) &:= a + \sum_{k=1}^N \left(b_k \diss^{\testx^T_k}(x) + c_k\adiss^{\testx^T_k}(y)\right)
 \end{align}    
 is an inclusion function for $\barrier^{\testx^T_{[1;N]}}$.

%
\subsubsection*{Sampling-based computation of T-RBCs}

Using the $T$-truncated dominance functions and the hyper-rectangular partition ${[\underline{x}^i,\overline{x}^i]}_{i\in \mc{I}}$ as sample points, the optimization problem~\eqref{eq:OP} for computing T-RBCs can be reduced to:
\begin{align}\label{eq:rvsop}
\min_{p \in \mc{P}}&\;\;\;\;\mathsf{Loss}(p)\nonumber\\
   \mbox{ s.t. }
     &\;\;\widehat{\barrier}^{\testx^T_{[1;N]}}(p,\overline{x}^i+\eps_T\vect{1}_n,\underline{x}^i + \eps_T\vect{1}_n) \leq 0,  &&\forall i \in \mc{I}_{0},\nonumber\\
     &\;\;\widehat{\barrier}^{\testx^T_{[1;N]}}(p,\underline{x}^i, \overline{x}^i)  - \sum_{k=1}^N\frac{b_k + c_k}{T+1} > 0,  &&\forall i\in \mc{I}_{u},\nonumber\\
     &\;\;b_k,c_k \geq 0,\;\;\forall k \in [1;N] 
     \tag{\textbf{R-SpOP}}
 \end{align}
 
We will show that any solution to the sampling-based problem~\eqref{eq:rvsop} yields a T-RBC for the monotone system $\mS$, and therefore guarantees its robust safety. Compared to~\eqref{eq:OP}, the primary advantage of the new formulation~\eqref{eq:rvsop} is that condition~\eqref{eq:trbc} is enforced solely through truncated trajectories and in finitely many sample points of $\mX$. 

 \begin{theorem}[Sampling-based T-RBC]\label{thm:rvSpOP}
Consider an SM system $\mS = (\mX,\mX_0,\mW,f)$ with the unsafe set $\mX_u\subseteq \mX$ and an unknown state transition map $f$ satisfying Assumptions \ref{as:test_verification}, \ref{as:lipschitz}, \ref{as:convergence}, and \ref{as:compact}. Let $\{[\underline{x}^i,\overline{x}^i]\}_{i\in \mathcal{I}}$ be a hyper-rectangular partitioning of $\mX$ satisfying~\eqref{eq:indices}. Given a finite time horizon $T>0$, if there is $p^* \in \mc{P}$ that solves the optimization problem~\eqref{eq:rvsop}, then $\mS$ is robustly safe with respect $\mX_u$.
 \end{theorem}
 \paragraph*{Proof: }See Appendix\ref{ap:thmrvSpOP}.
 $\hfill\blacksquare$
\smallskip
\begin{remark}[Comparison with the literature]
Theorem~\ref{thm:rvSpOP} provides a data-driven approach for guaranteeing robust safety and constructing robust invariant sets for monotone systems without restricting the search to specific parameterizations.
This contrasts with existing methods, which restrict the search to particular classes of sets, such as sublevel sets of SOS polynomials in~\cite{luppi2024data}, lower-closed sets in~\cite{MA-AS:24}, or hyper-rectangles in~\cite{abate2022robustly}.
Thus, Theorem~\ref{thm:rvSpOP} enables construction of robust invariant sets of arbitrary shape using trajectory data.

\end{remark}
%
\smallskip
\subsubsection*{Sampling-based computation of T-CBCs}

Similarly, using the truncated controlled dominance functions and the hyper-rectangular partition ${[\underline{x}^i,\overline{x}^i]}_{i\in \mc{I}}$ as sample points, the optimization problem~\eqref{eq:OP} for computing T-CBCs can be reduces to:
  \begin{align}\label{eq:cssop}
\min_{p \in \mc{P}}&\;\;\;\;\mathsf{Loss}(p)\nonumber\\
   \mbox{ s.t. }
     &\;\;\widehat{\barrier}^{\testx^T_{[1;N]}}(p,\overline{x}^i,\underline{x}^i) \leq 0,  &&\forall i \in \mc{I}_{0},\nonumber\\
     &\;\;\widehat{\barrier}^{\testx^T_{[1;N]}}(p,\underline{x}^i, \overline{x}^i) > 0,  &&\forall i\in \mc{I}_{u},\nonumber\\
     &\;\;\max_{k^q \in K^q}\ctrl_{k^q}(\underline{x}^i) \le \min_{k^p \in K^p} \ctrl_{k^p}(\overline{x}^i), &&\forall i\in \mathcal{I}, \nonumber\\
     &\;\;b_k,c_k \geq 0,\;\;\forall k \in [1;N];\tag{\textbf{C-SpOP}}
 \end{align}
with $K^p := \{k \in [1;N] \mid b_k > 0\}$ and $K^q := \{k \in [1;N] \mid c_k > 0\}$.
We show that any solution to the sampling-based problem~\eqref{eq:cssop} yields a T-CBC for the SIM system $\mS'$ and therefore shows that its safety controllable.  
 \begin{theorem}[Sampling-based T-CBC]\label{thm:csSpOP}
Consider an SIM system $\mS' = (\mX,\mX_0,\mU,f)$ with the unsafe set $\mX_u\subseteq \mX$ and an unknown state transition map $f$ satisfying Assumptions \ref{as:test_control}, \ref{as:finalxT}, and \ref{as:compact}. Let $\{[\underline{x}^i,\overline{x}^i]\}_{i\in \mathcal{I}}$ be a hyper-rectangular partitioning  of $\mX$ satisfying~\eqref{eq:indices}. Given a finite time horizon $T>0$, if there is $p^* \in \mc{P}$ that solves the optimization problem~\eqref{eq:cssop}, then $\mS'$ with any feedback controller $\pi:\mX\to \mU$ satisfying
\begin{align*}
    \pi(x) \in \bigcap_{k^p \in K^p}[\{\ctrl_{k^p}(\underline{x}^i)\}]_{\downarrow}\bigcap_{k^q \in K^q}[\{\ctrl_{k^q}(\overline{x}^i)\}]_{\uparrow},\;\; x\in [\underline{x}^i,\overline{x}^i],
\end{align*}
is controlled safe with respect to $\mX_u$.
 \end{theorem}
 \paragraph*{Proof: }
    See Appendix\ref{ap:thmcsSpOP}.
 $\hfill\blacksquare$
\smallskip
\begin{remark}\;
\begin{enumerate}
    \item \textit{(Comparison with the literature).}
    Theorem~\ref{thm:csSpOP} provides a systematic framework for computing controlled invariant sets of monotone systems. In comparison, \cite[Proposition 7]{MA-AS:24} constructs a lower-closed controlled invariant set under the weaker assumption that $\testx(T) \in [\bigcup_{t = 0}^{T-1}\testx(t)]_{\downarrow}$, by taking the union of the sets $[\testx(t)]_{\downarrow}$. In contrast, our method leverages dominance functions to explicitly learn the geometry of the controlled invariant set, enabling constructions that can better capture and adapt to the desired safety specification.

    \item \textit{(Sparsity Promoting Loss).} While it is desirable to promote sparsity in the parameter vector $p \in \mc{P}$, which contains the coefficients $a$, $b_k$, and $c_k$ of the T-CBC, the sparsity of these coefficients directly influences the admissible controller set in Theorem~\ref{thm:csSpOP}. In particular, increasing the number of nonzero coefficients typically imposes more constraints on the controller domain in Theorem~\ref{thm:csSpOP}.  This reveals an inherent trade-off between the expressive power of the T-CBC parameterization and the existence of feasible control actions.
    
    \item \textit{(Mixed Integer Linear formulation).} When the loss function is linear in $p$, the optimization problems~\eqref{eq:rvsop} and~\eqref{eq:cssop} can be reformulated as mixed-integer linear programs (MILPs) by introducing indicator constraints~\cite{PB-AL-AT-SW:15} to enforce $b_k = 0$ and $c_k = 0$ whenever required. This reformulation makes the optimization problems amenable to branch-and-bound techniques~\cite{DRM-SHJ-JJS-ECS:16}.
\end{enumerate}
\end{remark} 

\section{Numerical Experiments}
In this section, we verify safety of a 5-dimensional Lotka--Volterra mutualistic population dynamics model and synthesize a safe controller for a 2-dimensional traffic flow system\footnote{All simulations are performed on a PC with Ubuntu 20.04, Processor Intel i9-9900K$\times$16, and Memory RAM 32GB. We use GUROBI to solve the optimization problems. The code for these experiments is available at \url{https://github.com/HyConSys/Trajectory-basedSynthesisVerification}.}.
\begin{example}[Verification of Population Dynamics]\label{ex:nonlinear5D} 
We study the population evolution of mutualistic species using the Lotka--Volterra model presented in \cite{bullo2026contraction}. Given only bounded initial populations, we aim to formally verify that each species remains within a prescribed safe range over time.
We model this population evolution as a discrete-time system $\mS = (\mX, \mX_0, \mW, f)$, where the state space is $\mX = [0.1, 10]^5$, the set of initial conditions is $\mX_0 = [4, 6]^5$, and the disturbance set is empty ($\mW = \emptyset$). The transition map $f$ is given by
\begin{equation}\label{eq:exnonlinear5D}
\begin{aligned}
    f(x) = x + \tau\diag{x}[Ax + r - &\diag{r/K}x]\\
A = \begin{bmatrix}
0.00 & 0.02 & 0.00 & 0.00 & 0.00 \\
0.01 & 0.00 & 0.00 & 0.02 & 0.02 \\
0.00 & 0.00 & 0.00 & 0.01 & 0.02 \\
0.00 & 0.02 & 0.02 & 0.00 & 0.00 \\
0.00 & 0.01 & 0.01 & 0.00 & 0.00
\end{bmatrix},
r = &\begin{bmatrix}
0.22 \\
0.29 \\
0.26 \\
0.25 \\
0.23
\end{bmatrix},
K = \begin{bmatrix}
3.81 \\
2.47 \\
4.23 \\
2.93 \\
4.89
\end{bmatrix},
\end{aligned}
\end{equation}
where $\diag{y}$ is a diagonal matrix with components of vector $y$ on its diagonal. The unsafe set is defined as $\mX_u = [0.1, 2]^5 \cup [8, 10]^5$. For $\tau\le 2.2$, system $\mS$ is monotone \cite{bullo2026contraction}; for this example, we choose $\tau = 0.2$. We assume that neither the transition map $f$ nor a simulator of $f$ is available, and that only two system trajectories can be collected (see Fig.~\ref{fig:ex1_traj}) with $T=400$. These trajectories start from $\testx_1^T(0) = (1.46,0.84, 0.67, 1.59,0.78)$ and $\testx_2^T(0) = (8.65,9.74,8.83,9.17,9.61)$, respectively.
\begin{figure}
    \centering
    \includegraphics[width=1\linewidth]{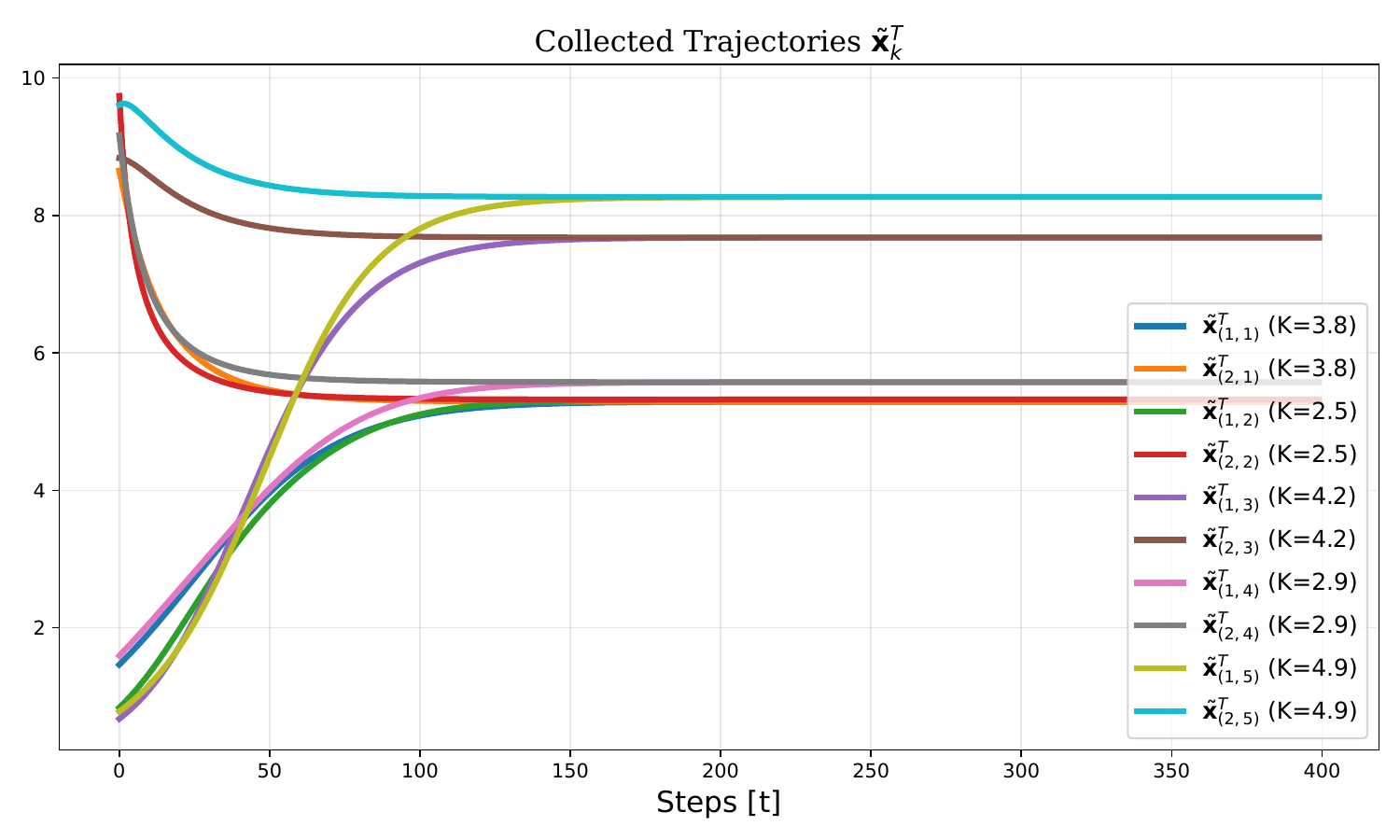}
    \caption{Example \ref{ex:nonlinear5D}. Collected test trajectories $\testx_1^T,\testx_2^T$ with $T = 400$ for safety verification; where $\testx^T_{(k,j)}$ is the $j-$th state component of the $k-$th collected trajectory.} 
    \label{fig:ex1_traj}
    \vspace{-0.5cm}
\end{figure}
Our goal is to verify the safety of $\mS$ using only these trajectories. By Theorem~\ref{thm:rvSpOP}, the search for a T-RBC can be formulated as the optimization problem~\eqref{eq:rvsop} with five decision variables $p := (a,b_1,b_2,c_1,c_2)$. We create a hyper-rectangular partition for $\mX$ such that$\norm{\underline{x}^i - \overline{x}^i}=0.5$. We solve~\eqref{eq:rvsop} in 7.2865 seconds, and construct the following T-RBC for the system:   
\begin{equation}\label{eq:ex_ver_barrier}
\begin{aligned}
    \barrier^{\testx^T_{[1;2]}}(x) = -0.069 + 0.066\adiss^{\testx_1^T}(x)
    + 0.151\diss^{\testx_2^T}(x).    
\end{aligned}    
\end{equation}
This example illustrates another benefit of our approach: it can perform safety verification with \textit{few} trajectories.
\end{example}
\bigskip
\begin{example}[Safe Traffic Control Synthesis]\label{ex:linear2D} We study the traffic flow dynamics between two road segments modeled by Daganzo’s cell transmission model \cite{como2017resilient}. We assume control over the amount of traffic entering each road segment through a saturation-based feedback controller.
We model this traffic control system using the discrete-time system $\mS = (\mX,\mX_0,\mU,f)$ where $\mX = [0,10]^2$, $\mX_0 = [4,6]^2$, $\mU = [0,10]\times[0.1,0.9]$, transition map $f$ defined by
\begin{equation}\label{eq:exlinear2D}
\begin{aligned}
 f_1(x,u) &= x_1 + \tau(u_1-\varphi_1(x_1))\\
 f_2(x,u) &= x_2 + \tau(u_2\mathbf{1}_{[0,x^{\max}_2]}(x_2)\varphi_1(x_1) - \varphi_2(x_2)),
\end{aligned}
\end{equation}
where $\tau$ is the step-size and $\mathbf{1}_{A}$ is the indicator function defined by $\mbf{1}_{A}(x) = \{1 \text{ if } x \in A, 0 \text{ if } x \notin A\}$ and $\varphi_j(x_j) = x^{\max}_j(1-\exp(-x_j))$ for $j \in [1;2]$. We assume that the unsafe set is given by $\mX_u = [0,1]^2\cup ([0,10]\times [9,10])\cup ([9,10]\times[0,10])$. One can show that this system is monotone for $\tau\le 0.01$~\cite[Lemma 4.29]{bullo2026contraction}.

We assume that neither the transition map $f$ nor a simulator of $f$ is available. Instead, we collect two trajectories: $\testx_1$, initialized at $\testx_1(0) = (9.5,9.9)$ under the monotone control policy $\ctrl_1 = (9,0.6)$, and $\testx_2$, initialized at $\testx_2(0) = (0.1,0.3)$ under the monotone control policy $\ctrl_2 = (9,0.5)$. These trajectories are collected over a horizon $T=1000$. 
Our goal is to design a safe controller that formally guarantees the safety of the system $\mS$ with respect to $\mX_u$. We create a hyper-rectangular partition for $\mX$ such that $\norm{\underline{x}^i - \overline{x}^i} = 1$. We solve the optimization problem in \eqref{eq:cssop} in 0.1956 seconds, and obtain the following T-CBC:
\begin{equation}\label{eq:exlinear_tcbc}
 \begin{aligned}
\barrier^{\testx^T_{[1;2]}}(x) = -0.115 + 1.073\diss^{\testx^T_1}(x)+0.408\adiss^{\testx^T_2}(x),
\end{aligned}   
\end{equation}
whose $0-$sublevel set is shown in Fig. \ref{fig:B_lin_ctrl_1}.
\begin{figure}
    \centering
    \includegraphics[width=1\linewidth]{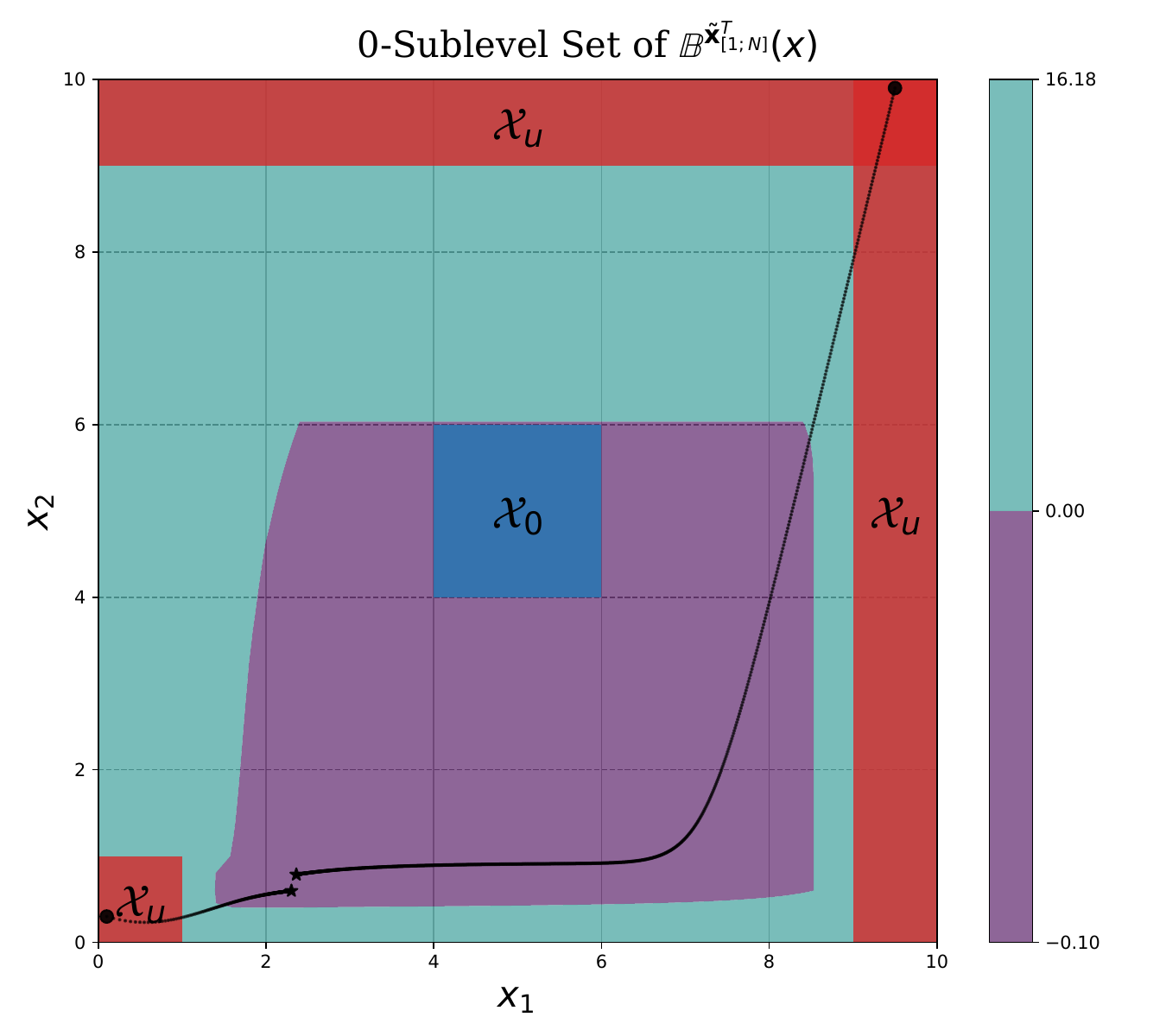}
    \vspace{-5mm}
    \caption{Example \ref{ex:linear2D}. The area in red is the unsafe set $\mU$, the area in blue is the initial set $\mX_0$. The two sequences show $\testx_1^T$ and $\testx_2^T$ starting at the points marked with black circles and ending at points marked with the stars. The area in purple is the $0-$sublevel set of $\barrier^{\testx^T_{[1;2]}}$ in \eqref{eq:exlinear_tcbc}.}
    \label{fig:B_lin_ctrl_1}
    \vspace{-5mm}
\end{figure}
Moreover, we obtain the following feedback controller
\begin{align*}
    \pi(x) &\in [\{\ctrl_1(\underline{x}^i)\}]_{\downarrow}\cap[\{\ctrl_2(\overline{x}^i)\}]_{\uparrow} =\{9\}\times[0.5,0.6]
\end{align*}
where $x \in [\underline{x}^i,\overline{x}^i]$. This means that we can apply a constant value of $\pi(x)$ for all $x \in \mX$ and the system renders safe with respect to unsafe set $\mX_u$. 
\end{example}

\section{Conclusion}\label{sec:conclusion}
In this work, we provided a trajectory-based framework to solve the safety problem for discrete-time monotone systems with an unknown transition map, in two cases: robust verification and control synthesis. To do so, we collect trajectory data to construct \emph{dominance functions} that are monotone and non-increasing along the system's evolution. We consider these functions as a basis to synthesize (control) barrier certificates. 
In future work, we will explore different compositions of dominance functions, such as nonlinear functions or multidimensional functions, for the construction of trajectory-based barriers. Moreover, we are interested in the computational traceability of these certificates for specifications more complex than safety, and in the case of general conic preorders describing the monotonicity of the system.

\bibliographystyle{unsrtnat}
\bibliography{ref.bib}  

\section*{APPENDIX}
\subsection{Proof of Proposition \ref{thm:traj_comp}}\label{ap:thmtraj_comp}
We prove this result by induction on $t$. For the base case, we have $\testx(0) = \hatx(0)$. For the inductive step, assume the claim holds for some $t$, meaning that 
\begin{align}\label{eq:induction}
     - L_wD_w\left(\textstyle\sum_{\tau = 0}^{t-1}L_x^\tau\right)\vect{1}_n\le \hatx(t)-\testx(t) \le  L_wD_w\left(\textstyle\sum_{\tau = 0}^{t-1}L_x^\tau\right)\vect{1}_n
\end{align}
We now show that it also holds for $t+1$. First, note that, $f$ is monotone. Therefore, by applying $f$ to the right hand side of~\eqref{eq:induction}, we get  $f(\hatx(t),\testw(t)) \le f\left(\testx(t) + L_w D_w\left(\textstyle\sum_{\tau = 0}^{t-1}L_x^\tau\right)\vect{1}_n,\testw(t)\right)$. This implies that 
\begin{align*}
    \hatx&(t+1) = f(\hatx(t),\hatw(t)) + f(\hatx(t),\testw(t)) - f(\hatx(t),\testw(t))\\
    &\leq f\left(\testx(t) + L_wD_w\left(\textstyle\sum_{\tau = 0}^{t-1}L_x^\tau\right)\vect{1}_n,\testw(t)\right)\\
    & \quad + f(\hatx(t),\hatw(t)) - f(\hatx(t),\testw(t)),
\end{align*}
Now, by Lipschitz continuity of $f$ with respect to $w$ we obtain,
\begin{align*}
    \hatx(t+1)
     \leq f\left(\testx(t) + L_wD_w\left(\textstyle\sum_{\tau = 0}^{t-1}L_x^\tau\right)\vect{1}_n,\testw(t)\right) + L_w D_w\vect{1}_n
\end{align*}
Using Lipschitz continuity of $f$ with respect to $x$, we get
\begin{align*}
 \hatx&(t+1) \\ 
    &\leq f(\testx(t),\testw(t)) + L_xL_wD_w\left(\textstyle\sum_{\tau = 0}^{t-1}L_x^\tau\right)\vect{1}_n + L_wD_w\vect{1}_n\\
    & = \testx(t+1) + L_wD_w\left(L_x\left(\textstyle\sum_{\tau = 0}^{t-1}L_x^\tau\right)\vect{1}_n + \vect{1}_n\right)\\
    & = \testx(t+1) + L_wD_w\left(\textstyle\sum_{\tau = 0}^{t}L_x^\tau\right)\vect{1}_n.
\end{align*}
Similarly, by applying the monotone $f$ to the right hand side of~\eqref{eq:induction}, we get $f(\hatx(t),\testw(t)) \ge f\left(\testx(t) - L_w D_w\left(\textstyle\sum_{\tau = 0}^{t-1}L_x^\tau\right)\vect{1}_n,\testw(t)\right)$. Thus,
\begin{align*}
    \hatx&(t+1) = f(\hatx(t),\hatw(t)) + f(\hatx(t),\testw(t)) - f(\hatx(t),\testw(t))\\
    & \geq f\left(\testx(t) - L_wD_w\left(\textstyle\sum_{\tau = 0}^{t-1}L_x^\tau\right)\vect{1}_n,\testw(t)\right)\\
    & \quad + f(\hatx(t),\hatw(t)) - f(\hatx(t),\testw(t))
 \end{align*}
 Now, by Lipschitz continuity of $f$ with respect to $w$, we get 
 \begin{align*}
     \hatx&(t+1) \\
     & \geq f\left(\testx(t) - L_wD_w\left(\textstyle\sum_{\tau = 0}^{t-1}L_x^\tau\right)\vect{1}_n,\testw(t)\right) - L_w D_w\vect{1}_n\\
    & \geq f\left(\testx(t) - L_wD_w\left(\textstyle\sum_{\tau = 0}^{t-1}L_x^\tau\right)\vect{1}_n,\testw(t)\right) - L_w D_w\vect{1}_n
    \end{align*}
    Additionally, by Lipschitz continuity of $f$ with respect to $x$,
    \begin{align*}
    \hatx&(t+1) \\
    & \geq f(\testx(t),\testw(t)) - L_x L_wD_w\left(\textstyle\sum_{\tau = 0}^{t-1}L_x^\tau\right)\vect{1}_n - L_wD_w\vect{1}_n\\
    & = \testx(t+1) - L_wD_w\left(L_x\left(\textstyle\sum_{\tau = 0}^{t-1}L_x^\tau\right)\vect{1}_n + \vect{1}_n\right)\\
    & = \testx(t+1) - L_wD_w\left(\textstyle\sum_{\tau = 0}^{t}L_x^\tau\right)\vect{1}_n.\qquad\qquad \qquad \quad \blacksquare
\end{align*}
\subsection{Proof of Theorem \ref{thm:a-disshatfcn}}\label{ap:thmadisshatfcn}

\emph{Properties of $\diss^{\testx}$.} For every $z\in \mX$, we define the set
\begin{align*}
 \mc{T}^{\testx}_{z}: = \{t \in \NN \mid z \leq \testx(t) + \lipconst{t-1}\vect{1}_n\}
\end{align*}
where $\lipconst{t}$ is as defined in~\eqref{eq:compact}. 

Regarding part~\ref{ph1}, consider $x \leq y$, our goal is to show that $\diss^{\testx}(x) \le \diss^{\testx}(y)$. To prove this, we consider two cases: (i) $\mc{T}^{\testx}_{y} \neq \emptyset$, and (ii) $\mc{T}^{\testx}_{y} = \emptyset$. For case (i), using the fact that $x\le y$, we get $x \leq y \leq \testx(t) + \lipconst{t-1}\vect{1}_n$, for every $t\in \mc{T}^{\testx}_{y}$. This implies that $\mc{T}^{\testx}_x \supseteq \mc{T}^{\testx}_{y}$ and since $\mc{T}^{\testx}_{y}\neq \emptyset$, we get that $\mc{T}^{\testx}_{x} \neq \emptyset$. On the other hand, we get
\begin{align*}
    t^{\le}(x) = \sup_{t\in \mc{T}^{\testx}_{x}} \{t\} \ge \sup_{t\in \mc{T}^{\testx}_{y}} \{t\} = t^{\le}(y) < \infty,
\end{align*}
where the first inequality holds because $\mc{T}^{\testx}_x \supseteq \mc{T}^{\testx}_{y}$ and the last strict inequality holds because $\mc{T}^{\testx}_{y} \neq \emptyset$. Now using the definition of robust upper dominance function $\diss^{\testx}$ in~\eqref{eq:disshatfcn}, 
\begin{align*}
    \diss^{\testx}(x)  = \tfrac{1}{t^{\le}(x)+1} \le \tfrac{1}{t^{\le}(y)+1} = \diss^{\testx}(y). 
\end{align*}
For case (ii) by definition of the robust upper dominance function $\diss^{\testx}$ in~\eqref{eq:disshatfcn}, we get that $\diss^{\testx}(y) = \alpha$. Since $\diss^{\testx}(z) \le \alpha$, for every $z\in \mX$, we get that $\diss^{\testx}(x)\le \diss^{\testx}(y)$. 

Regarding part~\ref{ph2}, we consider two cases: (i) $\mc{T}^{\testx}_x  \neq \emptyset$, and (ii) $\mc{T}^{\testx}_x  = \emptyset$. For case (i), assume that $\testw$ is the disturbance associated to the trajectory $\testx$. Then, for every $t\ge 0$, we have
\begin{align}
    f&\left(\testx(t)+\lipconst{t-1}\vect{1}_n,w\right) \le f(\testx(t),w) + L_x(\lipconst{t-1}\vect{1}_n) \nonumber\\ & \le f(\testx(t),\testw(t))
+ L_x(\lipconst{t-1}\vect{1}_n) + L_w D_w \vect{1}_n
\nonumber\\ & = \testx(t+1) + \lipconst{t}\vect{1}_n.\label{eq:boundP}
\end{align}
Since $\mc{T}^{\testx}_x  \neq \emptyset$, by combining \eqref{eq:Phattime} and \eqref{eq:disshatfcn}, we obtain that for every $w\in \mW$,
\begin{align*}
    \diss^{\testx}(f(x,w)) &= \inf_{t\ge 0}\left\{\tfrac{1}{t+1}\mid f(x,w) \leq \testx(t) + \lipconst{t-1}\vect{1}_n\right\}\\
    &\le \inf_{t\ge 0}\left\{\tfrac{1}{t+2}\mid f(x,w) \leq \testx(t+1) + \lipconst{t}\vect{1}_n\right\}\\
    &\leq \inf_{t\ge 0}\left\{\tfrac{1}{t+2}\mid f(x,w) \leq f(\testx(t) + \lipconst{t-1}\vect{1}_n,w)\right\}\\
    &\leq \inf_{t\ge 0}\left\{\tfrac{1}{t+2}\mid x \leq \testx(t) + \lipconst{t-1}\vect{1}_n\right\}
    \\
    &\leq \inf_{t\ge 0}\left\{\tfrac{1}{t+1}\mid x \leq \testx(t) + \lipconst{t-1}\vect{1}_n\right\} = \diss^{\testx}(x),
    \end{align*}
    where the first inequality holds because 
    \begin{multline*}
        \left\{\tfrac{1}{t+2}\mid f(x,w) \leq \testx(t+1) + \lipconst{t}\vect{1}_n\right\} \\ \subseteq \left\{\tfrac{1}{t+1}\mid f(x,w) \leq \testx(t) + \lipconst{t-1}\vect{1}_n\right\},
    \end{multline*} 
    the second inequality holds because of the bound in~\eqref{eq:boundP}, the third inequality holds by monotonicity of the transition map $f$ with respect to $x$, and the fourth inequality holds because 
    $\tfrac{1}{t+2}\le \tfrac{1}{t+1}$, for every $t\ge 0$. This means that $\diss^{\testx}(f(x,w))\leq \diss^{\testx}(x)$ for all $w \in \mW$. For case (ii), by definition of robust upper dominance function $\diss^{\testx}(x) = \alpha$ in~\eqref{eq:disshatfcn}, we get   $\diss^{\testx}(x) = \alpha$. This means that $\diss^{\testx}(f(x,w)) \leq \alpha = \diss^{\testx}(x)$, for every $w\in \mathcal{W}$. 
    
Regarding part~\ref{ph3}, for every $x \in \left(\diss^{\testx}\right)_{\leq c}$ and every $w \in \mW$, using part~\ref{ph2}, we have $\diss^{\testx}(f(x,w)) \leq \diss^{\testx}(x)\leq c$. This implies that $\left(\diss^{\testx}\right)_{\leq c}$ is a forward invariant set for the system $\mS$. 

\emph{Properties of $\adiss^{\testx}$}. For every $z \in \mX$, we define the set
\begin{align*}
 \mc{T}^{\testx}_{z}: = \{t \in \NN \mid z \geq \testx(t) - \lipconst{t-1}\vect{1}_n\}
\end{align*}
where $\lipconst{t}$ is as defined in~\eqref{eq:compact}. Now, we proceed using arguments similar to those used to show the properties of $\diss^{\testx}$.

Regarding part~\ref{ph1}, consider $x \leq y$, our goal is to show that $\adiss^{\testx}(x) \ge \adiss^{\testx}(y)$. To prove this, we consider two cases: (i) $\mc{T}^{\testx}_{x} \neq \emptyset$, and (ii) $\mc{T}^{\testx}_{x} = \emptyset$. For case (i), using the fact that $x\le y$, we get $\testx(t) - \lipconst{t-1}\vect{1}_n \leq x \leq y$, for every $t\in \mc{T}^{\testx}_{x}$. This implies that $\mc{T}^{\testx}_x \subseteq \mc{T}^{\testx}_{y}$ and since $\mc{T}^{\testx}_{x}\neq \emptyset$, we get that $\mc{T}^{\testx}_{y} \neq \emptyset$. On the other hand, we get
\begin{align*}
    t^{\ge}(x) = \sup_{t\in \mc{T}^{\testx}_{x}} \{t\} \le \sup_{t\in \mc{T}^{\testx}_{y}} \{t\} = t^{\ge}(y) < \infty,
\end{align*}
Using the definition of robust upper dominance function $\adiss^{\testx}$ in~\eqref{eq:disshatfcn}, we obtain the result. 
For case (ii), by definition of $\adiss^{\testx}$ in~\eqref{eq:adisshatfcn}, we get that $\adiss^{\testx}(x) = \alpha$. Since $\adiss^{\testx}(z) \le \alpha$, for every $z\in \mX$, we get that $\adiss^{\testx}(x)\ge \adiss^{\testx}(y)$. 

Regarding part~\ref{ph2}, we consider two cases: (i) $\mc{T}^{\testx}_x  \neq \emptyset$, and (ii) $\mc{T}^{\testx}_x  = \emptyset$. For case (i), assume that $\testw$ is the disturbance associated to the trajectory $\testx$. Then, for every $t\ge 0$, we have
\begin{align*}
    f&\left(\testx(t)-\lipconst{t-1}\vect{1}_n,w\right) \ge f(\testx(t),w) - L_x(\lipconst{t-1}\vect{1}_n) \\
    & \ge f(\testx(t),\testw(t))
- L_x(\lipconst{t-1}\vect{1}_n) - L_w D_w \vect{1}_n\\ 
& = \testx(t+1) - \lipconst{t}\vect{1}_n.
\end{align*}
Since $\mc{T}^{\testx}_x  \neq \emptyset$, by combining \eqref{eq:Qhattime} and \eqref{eq:adisshatfcn}, we obtain that for every $w\in \mW$,
\begin{align*}
    \adiss^{\testx}(f(x,w)) &= \inf_{t\ge 0}\left\{\tfrac{1}{t+1}\mid f(x,w) \geq \testx(t) - \lipconst{t-1}\vect{1}_n\right\}\\
    &\le \inf_{t\ge 0}\left\{\tfrac{1}{t+2}\mid f(x,w) \geq \testx(t+1) - \lipconst{t}\vect{1}_n\right\}\\
    &\leq \inf_{t\ge 0}\left\{\tfrac{1}{t+1}\mid x \geq \testx(t) - \lipconst{t-1}\vect{1}_n\right\} = \adiss^{\testx}(x).
    \end{align*}
    This means that $\adiss^{\testx}(f(x,w))\leq \adiss^{\testx}(x)$ for all $w \in \mW$. For case (ii), by definition $\adiss^{\testx}(x) = \alpha$ in~\eqref{eq:adisshatfcn}, we get   $\adiss^{\testx}(x) = \alpha$. This means that $\adiss^{\testx}(f(x,w)) \leq \alpha = \adiss^{\testx}(x)$, for all $w\in \mW$. 
    
Regarding part~\ref{ph3}, for every $x \in \left(\adiss^{\testx}\right)_{\leq c}$ and every $w \in \mW$, using part~\ref{ph2}, we have $\adiss^{\testx}(f(x,w)) \leq \adiss^{\testx}(x)\leq c$. This implies that $\left(\adiss^{\testx}\right)_{\leq c}$ is a forward invariant set for $\mS$. $\hfill \blacksquare$

\subsection{Proof of Theorem~\ref{thm:RBC-mon}}\label{ap:thm:RBC-mon}

Regarding~\eqref{thm:RBC-mon2} $\implies$\eqref{thm:RBC-mon1}, the proof follows from the fact that the function $\barrier(x)$ defined in~\eqref{eq:RBC-mon} is an RBC for $\mS$ and therefore the system $\mS$ is robustly safe with respect to $\mX_u$~\cite{Prajna2004}. 
Regarding~\eqref{thm:RBC-mon1} $\implies$\eqref{thm:RBC-mon2}, we prove that the function $\barrier(x)$ defined in~\eqref{eq:RBC-mon} is an RBC for the system $\mS$ by showing that it satisfies conditions~\eqref{eq:rbc1}-\eqref{eq:rbc3}. 

Let $x\in \mX_0$ and $\testx\in \Sigma$ be the trajectory of system $\mS$ starting from $\testx(0) = x$. By definition of robust upper dominance time~\eqref{eq:Qhattime}, we get  $t^{\le}(x) = \sup\{t\mid x\le \testx(t)\} \ge 0,$
where the last inequality holds because $\testx(0) = x$. Therefore, by definition of the robust upper dominance function in~\eqref{eq:adisshatfcn}, we have  $\diss^{\testx}(x) \le 1$. Similarly, by definition of robust lower dominance time~\eqref{eq:Qt}, we get 
$t^{\ge}(x) = \sup\{t\mid x\ge \testx(t)\} \ge 0,$
where the last inequality holds because $\hatx(0) = x$. Therefore, by definition of the robust lower dominance function in~\eqref{eq:adissfcn}, we have  $\adiss^{\testx}(x) \le 1$. This implies that $\max\{\diss^{\testx}(x),\adiss^{\testx}(x)\}\le~1$. As a result, we obtain
\begin{align*}
  \inf_{\hatx\in \Sigma}\left\{\max\{\diss^{\hatx}(x),\adiss^{\hatx}(x)\}\right\} \le \max\{\diss^{\testx}(x),\adiss^{\testx}(x)\} \le 1.
\end{align*}
Therefore $\barrier(x) \le 0$ and $\barrier$ satisfies condition~\eqref{eq:rbc1}. 

Let $x\in \mX_u$. Our goal is to show that $\barrier(x)>0$. We prove this by contradiction. Assume that $\barrier(x)\le 0$. This means that $\inf_{\hatx\in \Sigma}\left\{\max\{\diss^{\hatx}(x),\adiss^{\hatx}(x)\}\right\} \le 1$. By definition of robust upper and lower dominance functions $\diss^{\hatx}$ and $\adiss^{\hatx}$ in~\eqref{eq:adisshatfcn} and~\eqref{eq:adissfcn}, respectively, the $\inf$ is achieved, i.e., there should exists a trajectory $\testx$ of the system $\mS$ such that 
\begin{align*}
  \max\{\diss^{\testx}(x),\adiss^{\testx}(x)\} \le 1.
\end{align*}
This implies that $\diss^{\testx}(x)\leq 1$. By the definition of the robust upper dominance function in~\eqref{eq:adisshatfcn}, there exists $t\geq 0$ such that $x\leq \testx(t)$. Likewise, since $\adiss^{\testx}(x)\leq 1$, the definition of the robust lower dominance function in~\eqref{eq:adissfcn} guarantees the existence of $s\geq 0$ such that $x\geq \testx(s)$. Hence,
$$\testx(s)\leq x\leq \testx(t).$$
This contradicts the assumption that $\mX_u$ can be expressed as a union of upper-closed and lower-closed sets. Since $x\in\mX_u$, the point $x$ must belong to at least one such component set. Suppose first that $x$ belongs to a lower-closed set contained in $\mX_u$. Because $\testx(s)\leq x$ and the set is lower closed, it follows that $\testx(s)$ also belongs to that set, and therefore $\testx(s)\in\mX_u$. This contradicts the robust safety of $\mS$ with respect to $\mX_u$, since $\testx(s)$ is a state reached by the trajectory $\testx$.
An analogous argument applies if $x$ belongs to an upper-closed set contained in $\mX_u$. Indeed, since $x\leq \testx(t)$ and the set is upper closed, it follows that $\testx(t)$ must also belong to that set, and hence $\testx(t)\in\mX_u$, again contradicting the robust safety of $\mS$. This shows that $\barrier(x)>0$ and therefore $\barrier$ satisfies condition~\eqref{eq:rbc2}. 

Let $x\in \mX$, and let $\hatx$ be an arbitrary trajectory of the system $\mS$. By the dissipation properties of the robust dominance functions established in Theorem~\ref{thm:a-disshatfcn}, we have $\diss^{\hatx}(f(x))\leq \diss^{\hatx}(x)$ and $
\adiss^{\hatx}(f(x))\leq \adiss^{\hatx}(x)$. Consequently,
\begin{align*}
\max\left\{\diss^{\hatx}(f(x)),\adiss^{\hatx}(f(x))\right\}
\leq
\max\left\{\diss^{\hatx}(x),\adiss^{\hatx}(x)\right\}.
\end{align*}
Taking the $\inf$ over all trajectories $\hatx\in\Sigma$ on both sides yields
\begin{align*}
\barrier(f(x)) & = \inf_{\hatx\in\Sigma}\left\{
\max\left\{\diss^{\hatx}(f(x)),\adiss^{\hatx}(f(x))\right\}\right\}
\\ & \leq
\inf_{\hatx\in\Sigma}\left\{
\max\left\{\diss^{\hatx}(x),\adiss^{\hatx}(x)\right\}\right\} = \barrier(x).
\end{align*}
Therefore, $\barrier$ satisfies condition~\eqref{eq:rbc3}. Thus, $\barrier$ is an RBC for $\mS$, and consequently $\mS$ is robustly safe with respect to $\mX_u$. $\hfill\blacksquare$

\subsection{Proof of Theorem \ref{thm:a-dissfcn}}\label{ap:thmadissfcn}
For every $z\in \mX$, we define
\begin{align*}
 \mc{T}^{\testx}_z:=\{t \in \NN \mid z \leq \testx(t)\}.   
\end{align*}
Regarding part~\ref{p1}, consider $x \leq y $. Our goal is to show that $\diss^{\testx}_{\ctrl}(x)\le \diss^{\testx}_{\ctrl}(y)$. To prove this, we consider two cases: (i) $ \mc{T}^{\testx}_y \neq \emptyset$, and (ii) $ \mc{T}^{\testx}_y =\emptyset$. For case (i), using the fact that $x\le y$, we get $x \leq y \leq \testx(t)$, for every $t\in \mc{T}^{\testx}_{y}$. This implies that $\mc{T}^{\testx}_x \supseteq \mc{T}^{\testx}_{y}$ and since $\mc{T}^{\testx}_{y}\neq \emptyset$, we get that $\mc{T}^{\testx}_{x} \neq \emptyset$. On the other hand, we get
\begin{align*}
    t_{\ctrl}^{\le}(x) = \sup_{t\in \mc{T}^{\testx}_{x}} \{t\} \ge \sup_{t\in \mc{T}^{\testx}_{y}} \{t\} = t_{\ctrl}^{\le}(y) < \infty,
\end{align*}
where the first inequality holds because $\mc{T}^{\testx}_x \supseteq \mc{T}^{\testx}_{y}$ and the last strict inequality holds because $\mc{T}^{\testx}_{y} \neq \emptyset$. Now using the definition of controlled upper dominance function $\diss_{\ctrl}^{\testx}$ in~\eqref{eq:dissfcn}, 
\begin{align*}
    \diss_{\ctrl}^{\testx}(x)  = \tfrac{1}{t_{\ctrl}^{\le}(x)+1} \le \tfrac{1}{t_{\ctrl}^{\le}(y)+1} = \diss_{\ctrl}^{\testx}(y). 
\end{align*}
For case (ii) by definition of the robust upper dominance function $\diss_{\ctrl}^{\testx}$ in~\eqref{eq:dissfcn}, we get that $\diss_{\ctrl}^{\testx}(y) = \alpha$. Since $\diss^{\testx}(z) \le \alpha$, for every $z\in \mX$, we get that $\diss_{\ctrl}^{\testx}(x)\le \diss_{\ctrl}^{\testx}(y)$. 

Regarding part~\ref{p2}, we consider three cases: (i) $\mc{T}^{\testx}_x  = \emptyset$, and (ii) $\mc{T}^{\testx}_x  = \emptyset$. For case (i), by monotonicity of the system and controller $\ctrl$, for every $u \in [\{\ctrl(x)\}]_{\downarrow}$, we have
\begin{align}\label{eq:bound2}
    f(\testx(t),u)\leq f(\testx(t),\ctrl(\testx(t))) = \testx(t+1).
\end{align}
Since $\mc{T}^{\testx}_x  \neq \emptyset$, for every $u \in [\{\ctrl(x)\}]_{\downarrow}$, by combining \eqref{eq:Ptime} and \eqref{eq:dissfcn}, we obtain that
\begin{align*}
    \diss_{\ctrl}^{\testx}(f(x,u)) &= \inf_{t\ge 0}\left\{\tfrac{1}{t+1}\mid f(x,u) \leq \testx(t)\right\}\\
    &\le \inf_{t\ge 0}\left\{\tfrac{1}{t+2}\mid f(x,u) \leq \testx(t+1)\right\}\\
    &\leq \inf_{t\ge 0}\left\{\tfrac{1}{t+2}\mid f(x,u) \leq f(\testx(t),u)\right\}\\
    &\leq \inf_{t\ge 0}\left\{\tfrac{1}{t+2}\mid x \leq \testx(t)\right\}
    \\
    &\leq \inf_{t\ge 0}\left\{\tfrac{1}{t+1}\mid x \leq \testx(t)\right\} = \diss_{\ctrl}^{\testx}(x),
    \end{align*}
    where the first inequality holds because
    \begin{align*}
        \left\{\tfrac{1}{t+2}\mid f(x,u) \leq \testx(t+1) + \lipconst{t}\vect{1}_n\right\} \subseteq \left\{\tfrac{1}{t+1}\mid f(x,u) \leq \testx(t) + \lipconst{t-1}\vect{1}_n\right\},
    \end{align*} 
    the second inequality holds because of the bound in~\eqref{eq:bound2}, the third inequality holds by monotonicity of the transition map $f$ with respect to $x$, and the fourth inequality holds because 
    $\tfrac{1}{t+2}\le \tfrac{1}{t+1}$, for every $t\ge 0$. This means that $\diss_{\ctrl}^{\testx}(f(x,w))\leq \diss_{\ctrl}^{\testx}(x)$ for all $u \in [\{\ctrl(x)\}]_{\downarrow}$.  For case (ii), by definition of controlled upper dominance function $\diss_{\ctrl}^{\testx}(x) = \alpha$ in~\eqref{eq:dissfcn}, we get $\diss_{\ctrl}^{\testx}(x) = \alpha$. This means that $\diss_{\ctrl}^{\testx}(f(x,u)) \leq \alpha = \diss_{\ctrl}^{\testx}(x)$, for every $u \in [\{\ctrl(x)\}]_{\downarrow}$.

Regarding part~\ref{p3}, for every $x \in \left(\diss^{\testx}{\ctrl}\right){\leq c}$ and every $u \in [{\ctrl(x)}]{\downarrow}$, using part~\ref{p2}, $\diss{\ctrl}^{\testx}(f(x,u)) \leq \diss_{\ctrl}^{\testx}(x)\leq c$. Thus, $\left(\diss_{\ctrl}^{\testx}\right)_{\leq c}$ is forward invariant for $\mS$.

The proof for $\adiss_{\ctrl}^{\testx}$ follows \textit{mutatis mutandis}. $\hfill \blacksquare$

\subsection{Proof of Theorem \ref{thm:rvPQT}}\label{ap:rvthmPQT}
We prove properties \ref{p1-trunc-rv} and \ref{p2-trunc-rv} for $\diss^{\testx^T}$. Regarding \eqref{p1-trunc-rv}, consider $x,y \in \mX$ such that $x\leq y$. If $\diss^{\testx^T}(y) = \alpha$, then, $\diss^{\testx^T}(x) \leq \alpha = \diss^{\testx^T}(y)$. If $\diss^{\testx^T}(y)\neq \alpha$, then, there exists $t \in [0;T]$ such that $\diss^{\testx^T}(y) = \frac{1}{t^{\leq_T}(y) +1}$. By assumption, $x -\eps_T\vect{1}_n \leq y - \eps_T \vect{1}_n$. This leads to 
\begin{align*}
t^{\leq_T}(x) &= \max_{t \in [0;T]}\left\{t \mid x - \eps_T\vect{1}_n \leq \testx(t) + \lipconst{t-1}\vect{1}_n\right\}\\
&\geq \max_{t \in [0;T]}\left\{t \mid y - \eps_T\vect{1}_n \leq \testx(t) + \lipconst{t-1}\vect{1}_n\right\} = t^{\leq_T}(y)   
\end{align*}
Thus, $\diss^{\testx^T}(x)\leq \diss^{\testx^T}(y)$. 

Regarding~\eqref{p2-trunc-rv}, we first show $\diss^{\testx}(x) \leq \diss^{\testx^T}(x + \eps_T\vect{1}_n)$, for every $x\in \mX$. We consider two cases: (i) there exists $t \in [0;T]$ such that $x \leq \testx(t)+\lipconst{t-1}\vect{1}_n$, and (ii) there does not exist a $t \in [0;T]$ such that $x \leq \testx(t)+\lipconst{t-1}\vect{1}_n$. For case (i), we have 
\begin{align*}
    &t^{\leq}(x) =  \sup\left\{t\mid x \leq \testx(t) + \lipconst{t-1}\vect{1}_n\right\}\\
    &= \sup\left\{t\mid (x + \eps_T\vect{1}_n) - \eps_n\vect{1}_n \leq \testx(t) + \lipconst{t-1}\vect{1}_n\right\}\\
    &\geq \max_{t \in [0;T]}\left\{t\mid (x + \eps_T\vect{1}_n) - \eps_n\vect{1}_n \leq \testx(t) + \lipconst{t-1}\vect{1}_n\right\}\\
    &= t^{\leq_T}(x + \eps_T\vect{1}_n).
\end{align*}
Hence, $\diss^{\testx}(x) \leq \diss^{\testx^T}(x+\eps_T\vect{1}_n)$. For case (ii), by definition, we get $\diss^{\testx^T}(x + \eps_T\vect{1}_n) =\alpha$. This implies that $\diss^{\testx}(x) \leq \alpha = \diss^{\testx^T}(x + \eps_T\vect{1}_n)$. Now, we show $\diss^{\testx^T}(x) - \frac{1}{T+1}\leq \diss^{\testx}(x)$. We consider three cases: (i) $t^{\leq}(x) > T$, (ii) $t^{\leq}(x) \in [0;T]$, and (iii) $t^{\leq}(x) = \emptyset$. For case (i), we have 
\begin{align*}
    \diss^{\testx^T}(x) - \tfrac{1}{T+1}\leq 0\leq \diss^{\testx}(x).
\end{align*}

For case (ii), we note that 
\begin{align*}
    t^{\leq}(x) & = \sup\left\{t\mid x \leq \testx(t) + \lipconst{t-1}\vect{1}_n\right\} \\ & = \max_{t\in [0;T]}\left\{t\mid x \leq \testx(t) + \lipconst{t-1}\vect{1}_n\right\}\\ &\le \max_{t\in [0;T]}\left\{t\mid x - \varepsilon \vect{1}_n \leq \testx(t) + \lipconst{t-1}\vect{1}_n\right\}= t^{\leq_T}(x),
\end{align*}
where the second equality hold using the fact that $t^{\leq}(x) \in [0;T]$ and the third inequality holds because 
\begin{align*}
     \left\{t\mid x - \varepsilon \vect{1}_n \leq \testx(t) + \lipconst{t-1}\vect{1}_n\right\}\subseteq \left\{t\mid x \leq \testx(t) + \lipconst{t-1}\vect{1}_n\right\}.
\end{align*}
This means that $t^{\leq}(x) \le t^{\leq_T}(x)\leq T$. As a result, we get 
\begin{align*}
  \diss^{\testx}(x)= \tfrac{1}{t^{\leq}(x)+1} \ge  \tfrac{1}{t^{\leq_T}(x)+1} = \diss^{\testx^T}(x) \geq \diss^{\testx^T}(x)- \tfrac{1}{T+1}.
\end{align*}
 For case (iii), we note that in this case $\diss^{\testx}(x) = \alpha$. Therefore, $\diss^{\testx^T}(x) - \frac{1}{T+1}\leq \diss^{\testx}(x) = \alpha$.
  
The proof for $\adiss^{\testx^T}$ follows \textit{mutatis mutandis}. $\hfill \blacksquare$

\subsection{Proof of Theorem \ref{thm:PQT}}\label{ap:thmPQT}
To show statement \ref{p1-trunc-cs} for $\diss^{\testx^T}$, for every $z\in \mX$, we define:
\[\mc{T}^{\testx^T}_z :=\{t \in [0;T-1] \mid z \leq \testx(t)\}\]
\textit{Monotonicity)} Let $x \leq y$ and consider two cases: i) $\mc{T}{\testx^T}_x \neq \emptyset$, if $\mc{T}{\testx^T}_y = \emptyset$, then $\diss^{\testx^T}(x) \leq \diss^{\testx^T}(y)$. If $\mc{T}{\testx^T}_y \neq \emptyset$, by assumption $x \le y \le \testx(t_{\ctrl}^{\le_T}(y))$, then $\mc{T}{\testx^T}_y\subseteq \mc{T}{\testx^T}_x$, and $\max\{\mc{T}{\testx^T}_y\} \le \max\{\mc{T}{\testx^T}_x\}$; thus $\diss^{\testx^T}(x)\le \diss^{\testx^T}(y)$. ii) $\diss^{\testx^T}(x) = \alpha$, by contradiction, let $\diss^{\testx^T}(y)<\alpha$, hence, there exists $t_{\ctrl}^{\le_T}(y) \in [0;T-1]$ such that $y\le \testx(t_{\ctrl}^{\le_T}(y))$, thus $x\le  \testx(t_{\ctrl}^{\le_T}(y))$, contradicting $\diss^{\testx^T}(x) = \alpha$. 

\textit{Dissipation)} Let us consider two cases: i) $t_{\ctrl}^{\le_T}(x) = \emptyset$, then the property holds for any value of $\diss^{\testx^T}(f(x,\ctrl(x)))$ if $f(x,\ctrl(x)) \in \mX$. ii) $t_{\ctrl}^{\le_T}(x) \in [0;T-1]$, we show that $t_{\ctrl}^{\le_T}(f(x,u)) \neq \emptyset$ for any $u \in [\{\ctrl(x)\}]_{\downarrow}$ and  $t_{\ctrl}^{\le_T}(x)\le t_{\ctrl}^{\le_T}(f(x,u))$ which directly implies the dissipation property. We analyze two cases: $t_{\ctrl}^{\le_T}(x) \in [0;T-2]$; hence, $x \le \testx(t_{\ctrl}^{\le_T}(x))$ implies $f(x,u) \le \testx(t_{\ctrl}^{\le_T}(x)+1)$ with $u \in [\{\ctrl(x)\}]_{\downarrow}$ thus $t_{\ctrl}^{\le_T}(f(x,u)) \in [1;T-1]$, in fact, by monotonicity of the system and the definition of $t_{\ctrl}^{\le_T}$, it must be true that $t_{\ctrl}^{\le_T}(x)+1 \le t_{\ctrl}^{\le_T}(f(x,u))$, hence, the result. In case $t_{\ctrl}^{\le_T}(x) = T-1$, we have $f(x,u) \le \testx^T(t_{\ctrl}^{\le_T}(x)+1) = \testx^T(T)$ which by Assumption \ref{as:finalxT} part \ref{as1:finalxTP}, is also bounded by $f(x,u) \le \testx(T)\le \testx(T-1)$, then $t_{\ctrl}^{\le_T}(f(x,u)) = T-1$ which is the largest value the function $t_{\ctrl}^{\le_T}$ can obtain; therefore, $t_{\ctrl}^{\le_T}(x) = t_{\ctrl}^{\le_T}(f(x,u))$ implying dissipation.

\textit{Invariance)} By the dissipation property, we claim that for all $x \in \diss^{\testx^T}_{\le c}$, $\diss^{\testx^T}(f(x,u)) \leq \diss^{\testx^T}(x)$ for any $u \in [\{\ctrl(x)\}]_{\downarrow}$. By induction, it follows that the trajectory $\mbf{x}\in \mX^{\omega}$ satisfies $\diss^{\testx^T}(\mbf{x})\le c$.
The proof for $\adiss^{\testx^T}$ follows \textit{mutatis mutandis}. $\hfill \blacksquare$

\subsection{Proof of Theorem \ref{thm:rvSpOP}}\label{ap:thmrvSpOP}
We show that if $p^* = p^* = (a^*,b^*_1,\ldots,b^*_N,c^*_1,\ldots,c^*_N)\in \mathcal{P}$ is a solution for \eqref{eq:rvsop} then  $\barrier^{\testx_{[1;N]}}(p^*,x)$ defined by  
\begin{align*}
    \barrier^{\testx_{[1;N]}}(p^*,x) = a^* + \sum_{k=1}^N \left(b^*_k \diss^{\testx_k}(x) + c^*_k\adiss^{\testx_k}(x)\right)
\end{align*}
is a T-RBC for system $\mS$. First, we show that if $b_k,c_k \geq 0$ for all $k \in [1;N]$, then $\widehat{\barrier}^{\testx^T_{[1;N]}}$ defined in~\eqref{eq:Tcbarrier1-data} is an inclusion function for $\barrier^{\testx^T_{[1;N]}}$ defined in~\eqref{eq:Tbarrier3}. For every $p\in \mathcal{P}$ and every $x\in [\underline{x},\overline{x}]$, we get
\begin{align*}
   \widehat{\barrier}^{\testx^T_{[1;N]}}(p,\underline{x},\overline{x}) & = a + \sum_{k=1}^N \left(b_k \diss^{\testx^T_k}(\underline{x}) + c_k\adiss^{\testx^T_k}(\overline{x})\right) \\ & \le a + \sum_{k=1}^N \left(b_k \diss^{\testx^T_k}(x) + c_k\adiss^{\testx^T_k}(x)\right).
\end{align*}
where the inequality follows from the facts that $b_k,c_k \geq 0$ for all $k \in [1;N]$, and that the truncated robust dominance functions $\diss^{\testx^T_k}$ and $-\adiss^{\testx^T_k}$ are monotone for every $k \in [1;N]$, by Theorem~\ref{thm:rvPQT} part~\ref{p1-trunc-rv}. Similarly, for every $p\in \mathcal{P}$ and every $x\in [\underline{x},\overline{x}]$, we get
\begin{align*}
   \widehat{\barrier}^{\testx^T_{[1;N]}}(p,\overline{x},\underline{x}) & = a + \sum_{k=1}^N \left(b_k \diss^{\testx_k}(\overline{x}) + c_k\adiss^{\testx_k}(\underline{x})\right) \\ & \ge a + \sum_{k=1}^N \left(b_k \diss^{\testx_k}(x) + c_k\adiss^{\testx_k}(x)\right),
\end{align*}
This means that $\widehat{\barrier}^{\testx^T_{[1;N]}}$ is an inclusion function for $\barrier^{\testx^T_{[1;N]}}$.

For every $x\in \mathcal{X}_0$, by~\eqref{eq:indices}, there exists $i\in \mathcal{I}_0$ such that $x \in [\underline{x}^i,\overline{x}^i]$. Thus, using the constraints on the initial set in~\eqref{eq:rvsop}, we get
\begin{align*}
    0\geq\widehat{\barrier}^{\testx^T_{[1;N]}}&(p^*,\overline{x}^i + \eps_T\vect{1}_n,\underline{x}^i +\eps_T\vect{1}_n)\\
    &\ge \barrier^{\testx^T_{[1;N]}}(p^*,x + \eps_T\vect{1}_n) \ge \barrier^{\testx_{[1;N]}}(p^*,x),
\end{align*}
where the second inequality holds because $\widehat{\barrier}^{\testx^T_{[1;N]}}$ is an inclusion function for $\barrier^{\testx^T_{[1;N]}}$ and the third inequality holds by Theorem~\ref{thm:rvPQT} part~\ref{p2-trunc-rv} and the fact that $b^*_k,c^*_k \geq 0$ for all $k \in [1;N]$ by the third constraint in~\eqref{eq:rvsop}. This means that $\barrier^{\testx_{[1;N]}}(p^*,x)$ satisfies condition~\eqref{eq:trbc1}. 

For every $x\in \mathcal{X}_u$, by~\eqref{eq:indices}, there exists $i\in \mathcal{I}_u$ such that $x \in [\underline{x}^i,\overline{x}^i]$. Thus, using the constraints on the unsafe set in~\eqref{eq:rvsop}, we get
\begin{align*}
    0 &< \widehat{\barrier}^{\testx^T_{[1;N]}}(p^*,\overline{x}^i, \underline{x}^i) - \sum_{k = 1}^N\tfrac{b_k^*+c_k^*}{T+1} \\ & \le \barrier^{\testx^T_{[1;N]}}(p^*,x) - \sum_{k = 1}^N\tfrac{b_k^*+c_k^*}{T+1}  \leq \barrier^{\testx_{[1;N]}}(p^*,x). 
\end{align*}
where the second inequality holds because $\widehat{\barrier}^{\testx^T_{[1;N]}}$ is an inclusion function for $\barrier^{\testx^T_{[1;N]}}$ and the third inequality holds by Theorem~\ref{thm:rvPQT} part~\ref{p2-trunc-rv} and the fact that $b^*_k,c^*_k \geq 0$ for all $k \in [1;N]$ by the third constraint in~\eqref{eq:rvsop}. This means that $\barrier^{\testx_{[1;N]}}(p^*,x)$ satisfies condition~\eqref{eq:trbc2}.

Finally, by the third constraint in~\eqref{eq:rvsop}, $b^*_k,c^*_k \geq 0$ for all $k \in [1;N]$. This means that $\barrier^{\testx_{[1;N]}}(p^*,x)$ satisfies conditions~\eqref{eq:trbc3} and is a T-RBC for $\mS$. Robust safety of $\mS$ with respect to $\mX_u$ then follows from Theorem~\ref{thm:robustsafety}.$\hfill\blacksquare$

\subsection{Proof of Theorem \ref{thm:csSpOP}}\label{ap:thmcsSpOP}
We show that if $p^* = (a^*,b^*_1,\ldots,b^*_N,c^*_1,\ldots,c^*_N)\in \mathcal{P}$ is a solution for \eqref{eq:cssop} then  $\barrier^{\testx^T_{[1;N]}}(p^*,x)$ defined by  
\begin{align}\label{eq:BB}
    \barrier^{\testx^T_{[1;N]}}(p^*,x) = a^* + \sum_{k=1}^N \left(b^*_k \diss_{\ctrl_k}^{\testx^T_k}(x) + c^*_k\adiss_{\ctrl_k}^{\testx^T_k}(x)\right)
\end{align}
is a CBC for system $\mS$ as in Definition~\ref{def:cbarrier}. with a corresponding safe feedback controller $\pi(x) \in \Pi(x)$ as defined in~\eqref{eq:Tcbarrier2}.

First, since $b_k^*,c_k^* \geq 0$ for all $k \in [1;N]$ in the \eqref{eq:cssop}, the same argument as in the proof of Theorem~\ref{thm:PQT} can be used to show that $\widehat{\barrier}^{\testx^T_{[1;N]}}(p^*,\underline{x},\overline{x})$ defined by 
\begin{align*}
    \widehat{\barrier}^{\testx^T_{[1;N]}}(p^*,\underline{x},\overline{x}) = a^* + \sum_{k=1}^N \left(b^*_k \diss_{\ctrl_k}^{\testx^T_k}(\underline{x}) + c^*_k\adiss_{\ctrl_k}^{\testx^T_k}(\overline{x})\right)
\end{align*}
is an inclusion function for $\barrier^{\testx^T_{[1;N]}}(p^*,x)$ defined in~\eqref{eq:BB}. 

For every $x\in \mX_0$, by~\eqref{eq:indices}, there exists $i\in \mathcal{I}_0$ such that $x \in [\underline{x}^i,\overline{x}^i]$. Thus, using the constraints on the initial set in~\eqref{eq:cssop}, we get
\begin{align*}
\barrier^{\testx^T_{[1;N]}}(p^*,x) \le \widehat{\barrier}^{\testx^T_{[1;N]}}(p^*,\overline{x}^i,\underline{x}^i) \le 0,
\end{align*}
where the first inequality holds by the fact that $\widehat{\barrier}^{\testx^T_{[1;N]}}(p^*,\underline{x},\overline{x})$ is an inclusion function for $\barrier^{\testx^T_{[1;N]}}(p^*,x)$. This means that $\barrier^{\testx^T_{[1;N]}}(p^*,x)$ satisfies condition~\eqref{eq:cbc1}. 

For every $x\in \mX_u$, by~\eqref{eq:indices}, there exists $i\in \mathcal{I}_u$ such that $x \in [\underline{x}^i,\overline{x}^i]$. Thus, using the constraints on the unsafe set in~\eqref{eq:cssop}, we get 
\begin{align*}
&0 <\widehat{\barrier}^{\testx^T_{[1;N]}}(p^*,\underline{x}^i,\overline{x}^i) \leq \barrier^{\testx^T_{[1;N]}}(p^*,x), 
\end{align*}
where the first inequality holds by the fact that $\widehat{\barrier}^{\testx^T_{[1;N]}}(p^*,\underline{x},\overline{x})$ is an inclusion function for $\barrier^{\testx^T_{[1;N]}}(p^*,x)$. This means that, $\barrier^{\testx^T_{[1;N]}}(p^*,x)$ satisfies condition~\eqref{eq:cbc2}. 

Let $x\in \mX$, then there exists $i\in \mathcal{I}$ such that $x \in [\underline{x}^i,\overline{x}^i]$. By the third condition in \eqref{eq:cssop}, we have 
\begin{align}\label{eq:inclusionPI}
\Pi(\underline{x}^i,\overline{x}^i) = \bigcap_{k^p \in K^p}[\{\ctrl_{k^p}(\underline{x}^i)\}]_{\downarrow}\bigcap_{k^q \in K^q}[\{\ctrl_{k^q}(\overline{x}^i)\}]_{\uparrow} \neq \emptyset.     
\end{align}
Since $\ctrl_{k}:\mX\to \mU$ is a monotone map, for every $k\in [1;N]$, we have $\ctrl_{k^p}(\underline{x}^i) \le \ctrl_{k^p}(x)$, for every $k^p\in K^p$. This implies
\begin{align*}
    [\{\ctrl_{k^p}(\underline{x}^i)\}]_{\downarrow} \subseteq [\{\ctrl_{k^p}(x)\}]_{\downarrow}, \qquad\mbox{ for all }k^p\in K^p.
\end{align*}
Similarly, by monotonicity of $\ctrl_{k}:\mX\to \mU$, for every $k\in [1;N]$, we have $\ctrl_{k^q}(x)\le \ctrl_{k^q}(\overline{x}^i)$, for every $k^q\in K^q$. This means that
\begin{align*}
    [\{\ctrl_{k^q}(\underline{x}^i)\}]_{\uparrow} \subseteq [\{\ctrl_{k^q}(x)\}]_{\uparrow},\qquad\mbox{ for all }k^q\in K^q.
\end{align*}
Combining the above two inclusions with~\eqref{eq:inclusionPI}, we get 
\begin{align*}
\Pi(\underline{x}^i,\overline{x}^i) & = \bigcap_{k^p \in K^p}[\{\ctrl_{k^p}(\underline{x}^i)\}]_{\downarrow}\bigcap_{k^q \in K^q}[\{\ctrl_{k^q}(\overline{x}^i)\}]_{\uparrow} \\ & \subseteq \bigcap_{k^p \in K^p}[\{\ctrl_{k^p}(x)\}]_{\downarrow}\bigcap_{k^q \in K^q}[\{\ctrl_{k^q}(x)\}]_{\uparrow} = \Pi(x).     
\end{align*}
Thus, since $\Pi(\underline{x}^i,\overline{x}^i)\neq \emptyset$, it follows that $\Pi(x)\neq \emptyset$. For every $x\in \mX$, we choose $u\in \Pi(x)$ and thus
\begin{align*}
   \barrier^{\testx^T_{[1;N]}}&(p^*,f(x,u))  \le a^* + \sum_{k=1}^N \left(b^*_k \diss_{\ctrl_k}^{\testx^T_k}(x) + c^*_k\adiss_{\ctrl_k}^{\testx^T_k}(f(x,u))\right)
   \\ & \le a^* + \sum_{k=1}^N \left(b^*_k \diss_{\ctrl_k}^{\testx^T_k}(x) + c^*_k\adiss_{\ctrl_k}^{\testx^T_k}(x)\right) = \barrier^{\testx^T_{[1;N]}}(p^*,x).
\end{align*}
where the first inequality holds because $b_k^* \geq 0$ for all $k \in [1;N]$ and $u\in [\{\ctrl_{k^p}(x)\}]_{\downarrow}$, and thus by Theorem~\ref{thm:a-dissfcn} part~\ref{p2}, 
\begin{align*}
   \diss_{\ctrl_{k^p}}^{\testx^T_{k^p}}(f(x,u)) \le \diss_{\ctrl_{k^p}}^{\testx^T_{k^p}}(x), \qquad\mbox{ for all }k^p\in K^p.
\end{align*}
and third inequality holds because $c_k^* \geq 0$ for all $k \in [1;N]$ and $u\in [\{\ctrl_{k^q}(x)\}]_{\downarrow}$, and thus by Theorem~\ref{thm:a-dissfcn} part~\ref{p2}, we get
\begin{align*}
   \adiss_{\ctrl_{k^q}}^{\testx^T_{k^q}}(f(x,u)) \le \adiss_{\ctrl_{k^q}}^{\testx^T_{k^q}}(x), \qquad\mbox{ for all }k^q\in K^q. \testx 
\end{align*}
This shows that $\barrier^{\testx^T_{[1;N]}}(p^*,x)$ satisfies condition~\eqref{eq:cbc3}. Thus $\mS$ is controlled safe with respect to $\mX_u$. $\hfill\blacksquare$
\end{document}